\documentclass[iop]{emulateapj}   

\usepackage{ulem}
\usepackage{epsfig}
\usepackage{color}
\usepackage{natbib}
\usepackage{enumerate}
\newcommand{\fesc}{\ifmmode{f_{\rm esc}}\else{$f_{\rm esc}$}\fi}
\newcommand{\fescs}{\ifmmode{f_{\rm esc}^\star}\else{$f_{\rm esc}^\star$}\fi}
\newcommand{\kms}{\ifmmode{{\;\rm km~s^{-1}}}\else{km~s$^{-1}$}\fi}
\newcommand{\fgas}{\ifmmode{{f_{\rm gas}}}\else{$f_{\rm gas}$}\fi}
\newcommand{\cubecm}{\ifmmode{{\rm cm^{-3}}}\else{cm$^{-3}$}\fi}
\newcommand{\ztwo}{\ifmmode{{\rm [Z_2/H]}}\else{[Z$_2$/H]}\fi}
\newcommand{\zthree}{\ifmmode{{\rm [Z_3/H]}}\else{[Z$_3$/H]}\fi}
\newcommand{\lsim}{\lower0.3em\hbox{$\,\buildrel <\over\sim\,$}}
\newcommand{\gsim}{\lower0.3em\hbox{$\,\buildrel >\over\sim\,$}}

\newcommand{\eavg}{\ifmmode{\langle E_\gamma \rangle}\else{$\langle E_\gamma \rangle$}\fi}

\newcommand{\moray}{{\sl Enzo+Moray}}
\newcommand{\Ms}{\ifmmode{\textrm{M}_\odot}\else{$M_\odot$}\fi}
\newcommand{\vrms}{\ifmmode{v_{\rm rms}}\else{$v_{\rm rms}$}\fi}

\newcommand{\hh}{H$_2$}

\newcommand{\tvir}{\ifmmode{T_{\rm{vir}}}\else{$T_{\rm{vir}}$}\fi}
\newcommand{\mvir}{\ifmmode{M_{\rm{vir}}}\else{$M_{\rm{vir}}$}\fi}
\newcommand{\rvir}{\ifmmode{r_{\rm{vir}}}\else{$r_{\rm{vir}}$}\fi}

\newcommand{\jj}{\ifmmode{J_{21}}\else{$J_{21}$}\fi}
\newcommand{\flw}{\ifmmode{F_{LW}}\else{$F_{LW}$}\fi}
\newcommand{\kph}{\ifmmode{k_{\rm ph}}\else{$k_{\rm ph}$}\fi}

\newcommand{\zsun}{\ifmmode{\rm\,Z_\odot}\else{$\rm\,Z_\odot$}\fi}

\newcommand{\hi}{H {\sc i}}
\newcommand{\hii}{H {\sc ii}}
\newcommand{\hei}{He {\sc i}}
\newcommand{\heii}{He {\sc ii}}

\newcommand\unit[1]{\; \textrm{#1}}

\begin{document}

\shorttitle{X-RAY HEATING FROM POP III BINARIES}
\shortauthors{XU ET AL.}

\title{Heating the IGM by X-rays from Population III Binaries in High
  Redshift Galaxies}


\author{Hao Xu\altaffilmark{1},
 Kyungjin Ahn\altaffilmark{2},
John H. Wise\altaffilmark{3},    
Michael L. Norman\altaffilmark{1},
and Brian W. O'Shea\altaffilmark{4}}
\altaffiltext{1}{Center for Astrophysics and Space Sciences,
  University of California, San Diego, 9500 Gilman Drive, La Jolla, CA
  92093; hxu@ucsd.edu, mlnorman@ucsd.edu}
\altaffiltext{2}{Department of Earth Science Education, Chosun University,
Gwangju 501-759, Korea; kjahn@chosun.ac.kr}  
\altaffiltext{3}{Center for Relativistic Astrophysics, School of
  Physics, Georgia Institute of Technology, 837 State Street, Atlanta,
  GA 30332; jwise@gatech.edu}
\altaffiltext{4}{Lyman Briggs College and Department of Physics and Astronomy, Michigan State University,
East Lansing, MI 48824; oshea@msu.edu}

\begin{abstract}

Due to their long mean free path, X-rays are expected to have an
important impact on cosmic reionization by heating and ionizing the intergalactic 
medium (IGM) on large scales, especially after simulations have suggested that
Population III (Pop III) stars may form in pairs at redshifts as high as 20 -
30. We use the Pop III distribution and evolution from a
self-consistent cosmological radiation hydrodynamic simulation of the
formation of the first galaxies and a simple Pop III X-ray binary
model to estimate their X-ray output in a high-density region larger
than 100 comoving (Mpc)$^3$.  We then combine three different methods
--- ray tracing, a one-zone model, and X-ray background modeling ---
to investigate the X-ray propagation, intensity distribution, and long-term 
effects on the IGM thermal and ionization state. The efficiency
and morphology of photoheating and photoionization are dependent on
the photon energies. The sub-keV X-rays only impact the IGM near the
sources, while the keV photons contribute significantly to the X-ray
background and heat and ionize the IGM smoothly.  The X-rays just
below 1 keV are most efficient in heating and ionizing the IGM.  We
find that the IGM might be heated to over 100 K by $z=10$ and the high-density 
source region might reach 10$^4$ K, limited by atomic
hydrogen cooling. This may be important for predicting the 21 cm
neutral hydrogen signals. On the other hand, the free electrons
from X-ray ionizations are not enough to contribute significantly to
the optical depth of the cosmic microwave background to the Thomson scattering.

\end{abstract}


\section{Introduction}
\label{introduction}

The appearance of the first luminous objects marks the end of the cosmic
dark ages after recombination and the beginning of the last cosmic phase
transition of reionization. Subsequently, the universe begins to be
heated and ionized by radiation from first stars and galaxies and
their descendants. Physical processes of the heating and ionization
before the universe fully ionized at z $\sim$ 6 are still not
completely understood. It is one of the most important problems in
astrophysics and cosmology in both theory and observation to obtain
the thermal and ionization histories of the universe during this
transition period.

The first generation stars, Population III (Pop III), form from
metal-free gas in dark matter halos with M $\ga$ 10$^{6}$ M$_{\odot}$
and have a large characteristic mass \citep[e.g.,][]{Abel02, Bromm02,
  OShea07a, Turk09, Greif12_P3Cluster}. Due to their high mass, they
have short lifetimes \citep{Schaerer02}, may go supernova
\citep[SN; e.g.,][]{Heger03}, and enrich their surrounding intergalactic
medium (IGM). Once the metallicity of the star-forming gas passes some critical
metallicity, $\sim$ 10$^{-6}$ Z$_{\odot}$ if dust cooling is efficient
\citep{Omukai05, Schneider06, Schneider10, Omukai10, Dopcke13} or
$\sim$ 10$^{-3.5}$ Z$_{\odot}$ otherwise \citep{Omukai00, Bromm01,
  Smith09, Meece14}, the gas can cool rapidly and lower its Jeans mass.  These
metal-enriched Population II (Pop II) stars have a lower
characteristic mass scale and most likely have an initial mass
function (IMF) that resembles the present-day one.

Pop III stars can self-regulate their formation through chemical and
radiative feedback.  The transition from Pop III to Pop II star
formation and the end of the massive Pop III formation are strongly
dependent on the metal enrichment from the Pop III SN remnants in the
future star-forming halos. Metal enrichment involves complex
interactions between SNe blastwaves, the IGM, halo mergers, and
cosmological accretion.  This topic has been extensively studied with
semi-analytic models \citep{Scannapieco03, Yoshida04, Tumlinson06,
  Salvadori07, Komiya10, Gomez12, Crosby13b}, post-processing of numerical simulations
\citep{Karlsson08, Trenti09}, and direct numerical simulations
\citep{Tornatore07, Ricotti08, Maio10_Pop32, Wise12a, Muratov12,
  Xu13}. Studies have suggested that Pop III stars may continue to
form at low redshifts to the end of reionization. For example,
\citet{Trenti09} suggested that Pop III stars may still form at the
late epoch of $z = 6$ in the under dense regions of the universe by
post-processing of cosmological simulations with blast wave
models. \citet{Muratov12} also showed that Pop III stars continue to
form until $z = 6$ using direct cosmological simulations.

Pop III star formation is also regulated by the Lyman-Werner (LW)
radiation between 11.2 and 13.6 eV that is mostly produced by Pop III
and II stars.  LW photons photodissociate \hh~by the Solomon process
and then suppress the formation of Pop III stars in low-mass halos
\citep{Machacek01,Yoshida03, Wise07_UVB, OShea08}. Though LW radiation
will not completely suppress Pop III formation, it delays the
formation of Pop III stars by increasing the mass of Pop III forming
halos \citep{OShea08, Xu13}, which has complicated consequences on the
later star formation of both Pop II and Pop III by changing the metal
enrichment and radiation feedback.  For example, whether or not the enriched
gas can be ejected from the host halos is dependent on the halo masses
\citep{Whalen08, Muratov13b}.  In addition, since LW photons have long
mean free paths in the neutral IGM, LW radiation has a significant
impact on Pop III formation in distant halos ($\sim$ 100 comoving Mpc) and
usually needs to be handled carefully \citep[e.g.][]{Haiman00, Wise05,
  Ahn09}.

Because Pop III star formation is easily impacted by the environment
due to chemical and radiative feedback from nearby star-forming halos,
it is necessary to follow both the Pop II and Pop III formation and
feedback during early galaxy formation over a large cosmic volume to
track the their interaction between star-forming halos. In
\citet{Xu13}, we performed a self-consistent cosmological radiation
hydrodynamics simulation of Pop III and Pop II formation and feedback
in a volume larger than 100 comoving Mpc$^3$. This cosmological
simulation includes a full primordial chemistry network, radiative
cooling from metal species, models for both Pop II and Pop III star formation and
their radiative, mechanical, and chemical feedback, allowing us to
obtain a complete formation history of Pop III stars in a
statistically complete volume of $\sim$140 comoving Mpc$^3$, in which
over 13,000 Pop III stars form.

In addition to the metal enrichment and LW radiation from Pop III
stars, heating and ionizing effects from their radiation in other bands are crucial
to modeling early structure formation of the universe \citep{Gnedin97,
  Gnedin00, Haiman00}.  The Lyman continuum radiation from Pop III
stars then affects the subsequent structure formation through heating
and ionizing the surrounding IGM \citep{Machacek01, Yoshida03,
  Wise08_Gal, OShea08}.  The effects of this ionizing UV radiation
from stellar sources on the first galaxies are well studied
\citep{Ricotti02a, Ricotti02b, Wise08_Gal, Greif10, Wise12b, Wise12a,
  Muratov13b}.  However, the global impact of UV radiation from Pop III stars on 
  IGM is currently under debate. While it is agreed that Pop III stars cannot 
  finish reionization alone, they may contribute significantly by ionizing the 
  universe up to $\sim$ 20\% \citep[e.g.,][]{Haiman06, Ahn12} 
  or may be negligible \citep[e.g.,][]{Sobacchi13} due to negative radiative and 
  mechanical feedback effects \citep[e.g.,][]{OShea08, Whalen08_SN, Tseliakhovich10}. 
 This question has been addressed more self-consistently than before by 
 simulations in a large ($> \sim$100 Mpc) box (to incorporate large-scale 
 radiative feedback) with embedded subgrid microphysics on Pop III star formation 
 \citep{Ahn12,Fialkov13}, but a conclusive answer is yet to 
 come with even more self-consistent treatment of both global and local feedback effects.
 X-ray radiation from Pop III (stars or remnants) might also serve as a significant source of heating and 
ionization of the IGM during reionization, mainly due to its high efficiency 
in penetrating the IGM with much longer mean free path than UV radiation 
and thus generating global X-ray background. 

Simple vanilla models of reionization predict Thomson scattering optical depth 
of the cosmic microwave background (CMB), which usually differs a lot from the 
observed value. While the observation by Planck is
$\tau_e$ = 0.089 $\pm$ 0.032 \citep{2013arXiv1303.5076P}, instantaneous ionization produces 
$\tau_e$ $\sim$ 0.048 if it occurs at z $\sim$ 7, when the universe seems to have finished reionization.
This suggests that reionization is an extended process, starting as early as $z \sim
15-20$, and it has become clear that the majority of ionizing photons
originate from stellar sources \citep[e.g.,][]{Dijkstra04, Fan06,
 Willott10, Zahn12, Bouwens12_Reion, Haardt12, Becker13}.   X-ray radiation with 
 its much longer mean free path 
 than UV radiation has been considered a good candidate for the pre-ionization and pre-heating 
 much earlier than $z \sim 6$ \citep[e.g.,][]{Oh01, Venkatesan01, Ricotti04, Ricotti05}. In case the 
 integrated electron abundance by UV sources is not sufficient to explain the observed high 
 value of $\tau_e$, X-ray sources may generate additional electrons during epoch of reionization (EoR) 
 to compensate for such deficiency. X-ray pre-heating might have significant impacts on the 21 cm 
 signatures of reionization at high redshifts \citep[e.g.,][]{Mesinger13, Fialkov14}.  
 Detecting these signals is a very hard task with the current radio facilities, 
 but recent 21 cm observation of EoR  by PAPER \citep{Parsons13} 
 has suggested that the IGM at $z=7.7$ may have been pre-heated by X-rays.
Two major candidates of X-ray sources are active galactic nuclei (AGNs) and X-ray binaries (XRBs) of
metal-enriched stars. However, these sources appear in the universe late
($z \la 10$) and might be too weak to have an important impact on the
reionization history. In addition to these two candidates, Pop III 
stars and their remnants are possible candidates for strong X-ray emission.  
Pop III in the approximate mass range $40-140 \Ms$ and $>260 \Ms$ may 
directly collapse to form black holes \citep[BHs;][]{Heger03}. Any strong accretion onto
these massive Pop III seeding BHs would lead to X-ray radiation at high
redshifts \citep{Kuhlen05, Alvarez09, Tanaka12}.
Supporting this additional avenue of radiation sources during
reionization, recent cosmological simulations \citep{Turk09, Stacy10,
  Stacy13} have found that metal-free star-forming clouds might
fragment to form binaries in a non-negligible fraction of Pop III star-forming 
events, which are promising X-ray sources at high redshifts.

The emerging X-rays from these binaries might be excellent sources of
IGM pre-heating and pre-ionization for several reasons. They form at
very high redshifts, so there is ample time for X-rays to heat and
ionize the IGM, which is important because the ionization and heating
timescales can be a sizable fraction of the Hubble time. A top-heavy
IMF favors BH formation on the order of tens of solar masses, whose
luminosity are much higher than the late-type solar mass binaries, and
their radiation is strong at $\sim$ 1 keV, which aids in escaping deep into
the IGM, heating and partially ionizing the neutral gas in its path.
X-ray radiation from Pop III binaries has been suggested to produce a
pre-heated IGM \citep[e.g.,][] {Mirabel11, Haiman11, Mesinger13}and, 
more relevant to reionization, may partially ionize the IGM in
large volumes \citep[e.g.,][]{Ostriker96, Pritchard07}.

X-ray pre-heating and pre-ionization on large scales might also be
important for the later structure formation \citep{Tanaka12}.  For example, at high
redshift, heating from the CMB may limit the radiative cooling, and
thus increasing the Jeans mass, resulting in an IMF that also favors
massive star formation for Pop II stars \citep{Larson05,
  Smith09}. X-rays may also play the same role to heat the IGM and
change the IMF of the Pop II stars.  The impact of X-rays from Pop III
binaries has been tested by some recent cosmological simulations.
By considering X-ray feedback from Pop III binaries in a 1 Mpc$^3$
volume simulation, \citet{Jeon13} found that although no strong effects
on star formation history from the X-ray feedback were found, the preheating of
the IGM X-rays may lead to a suppression of small-scale structures and
lower the recombination rate in the IGM, which could accelerate the
reionization process.
 
X-rays from Pop III binaries are believed to be important for
reionization, and cosmic structure formation, and have been studied by
semi-analytic models \citep[e.g.][]{Power09, Mirabel11, Mesinger13}
and small-volume cosmological simulations \citep{Jeon13}.  In this
paper, we focus on using our results of Pop III formation in high-redshift 
galaxies in a large survey volume of over 100 comoving
Mpc$^{3}$ to understand the production and propagation of X-rays from
Pop III binaries, the corresponding X-ray background, and their impact
on the thermal and ionization state of the IGM.  This paper is
organized as follows.  We first describe our simulation and XRB 
model in Section 2. Then in Section 3, we present evolution and
distribution of X-ray emissions from Pop III binaries.  We study the
X-ray propagation, heating, and ionizing in the nearby IGM using ray tracing 
in Section 4.  In Section 5, we present our model of the X-ray background and our
estimations of the background and their effects on the IGM. Finally,
we conclude this study with a summary and a discussion of X-ray
background from Pop III binaries in the early universe, their IGM
heating and ionization, and possible observable effects in Section 6.


\section{Simulation and X-ray Binary model}
\label{simulation}

\subsection{Radiation Hydrodynamics Simulation: ``Rarepeak''}
We further analyze the simulation in \citet{Xu13} to study possible
X-ray radiation from Pop III binaries in the high-density
$\sim$ 138 comoving Mpc$^3$ survey volume.  The simulation is performed
using the adaptive mesh refinement (AMR) cosmological hydrodynamics
code Enzo \citep{Bryan13}.  The adaptive ray-tracing module
\moray~\citep{Wise11} is used for the radiation transfer of ionizing
radiation, which is coupled to the hydrodynamics and chemistry in
Enzo.

The initial conditions for the simulation is generated using
\textsc{Music} \citep{Hahn11_MUSIC} with second-order Lagrangian
perturbations at $z=99$.  We use the cosmological parameters from the
7-year WMAP $\Lambda$CDM+SZ+LENS best fit \citep{Komatsu11}:
$\Omega_{M}=0.266$, $\Omega_{\Lambda} = 0.734$, $\Omega_{b}=0.0449$,
$h=0.71$, $\sigma_{8}=0.81$, and $n=0.963$.  We use a hydrogen mass
fraction $X = 0.76$.  We simulate a comoving volume of (40 Mpc)$^3$
that has a $512^3$ root grid resolution and three levels of static
nested grids centered on a high-density region. We first run a 512$^3$ 
N-body-only simulation to $z=6$. Then we select the 
Lagrangian volume (a single rectangular box) around two $\sim 3 \times
10^{10} $ $\Ms$ halos at $z=6$ and re-initialize the simulation, having the Lagrangian
volume at the center, with three
more static nested grids to have an effective resolution of 4096$^3$ and an
effective dark matter mass resolution of $2.9 \times 10^4$ $\Ms$ inside
the highest nested grid, which just covers the Lagrangian volume,
with a comoving volume of $5.2 \times 7.0
\times 8.3\, \mathrm{Mpc}^3$ ($\sim$ 300 Mpc$^3$).  During the course of the
simulation, we allow a maximum refinement level $l=12$, resulting in a
maximal resolution of 19 comoving pc.  The refinement criteria
employed are the same as in \citet{Wise12a}.  The refinements
higher than the static nested grids are only allowed in a sub-volume,  
which adjusts its size during the simulation to contain only the 
highest-resolution dark matter particles, of the highest static nested grid.
The highly refined region, covering the Lagrangian volume of the two massive halos at $z=6$, 
has a comoving volume of $3.8 \times 5.4 \times 6.6$ 
Mpc$^3$ ($\sim$138 Mpc$^3$) 
at $z=15$, which represents a 3.5$\sigma$ density peak.  We call this 
well-resolved volume, which is also our survey volume, the Rarepeak in this
study and related papers \citep{Chen13, Ahn13}.  At this time, the
simulation has more than ten thousand Pop III stars and remnants
distributed over three thousand halos, most of them are more massive
than 10$^7$ M$_{\odot}$.  The simulation has 1.3 billion computational
cells in the refined region and consumed more than 10 million CPU hours 
on the Kraken system at NICS and Blue Waters system at NCSA.

\subsection{Star Formation and Feedback}

Both Pop II and Pop III stars are allowed to form inside the survey
volume, and we distinguish them by the total metallicity of the
densest star-forming cell. Pop III stars are formed if $[\textrm{Z/H}]
< -4$, and Pop II stars are formed otherwise. We use the same star
formation models and most of the parameters as in \citet{Wise12a}, as
well as feedback models. For the initial mass of Pop III stars, we
randomly sample from an IMF with a functional form:
\begin{equation}
f(\log M)dM=M^{-1.3}\exp\left[-\left(\frac{M_{\rm
        char}}{M}\right)^{1.6}\right]\, dM
\end{equation}
which behaves as a Salpeter IMF above the characteristic mass, M$_{\rm
  char}$, but is exponentially cutoff below that mass
\citep{Chabrier03}. Here, we use a characteristic mass of 40 M$_\odot$
for the Pop III IMF, which agrees with the latest results of Pop III
formation simulations \citep[e.g.,][]{Turk09, Greif12_P3Cluster}. For
the details of the star formation and stellar feedback schemes used,
refer to Sections 2.2 and 2.3 of \citet{Wise12a}.

The simulation performs ray tracing to calculate the propagation of UV
\hi~ionizing radiation, but to study the X-ray radiation transport and
associated photoheating and photoionization effects from Pop III
binaries, we post-process the data sets with \moray~to calculate the
X-ray radiation transport, which also includes the ionization of
\hei~and \heii.  We consider secondary ionizations and heating by
X-ray photons, using the fitting formula from \citet{Shull85}. Details
of the implementation can be found in \citet{Wise11}.

\subsection{X-Ray Model of Pop III Binaries}
\label{sec:xray_model}
  
We first need to use our Pop III distribution from simulation to estimate 
the X-ray radiation from Pop III binaries.
Since the occurrence fraction, properties, and evolution of Pop III
binaries are not yet well constrained \citep[e.g.,][]{Stacy13}, we build a simple model,
which takes advantage of solid information on Pop III population and distribution 
and ignores the details of Pop III binary formation and evolution, to estimate the 
X-ray luminosities from Pop III binaries in Rarepeak. We ignore the Pop III 
star initial mass, when we set the chance of a Pop III becoming binary 
and the evolution of the companion star.
We assume that there is $\alpha$ ($0 \le\alpha\le 1$) chance
that a Pop III star becomes an XRB. We also assume that one 
companion directly collapses into a BH with a
mass of tens of \Ms~without a supernova \citep{Fryer99}, and the
remaining mass exists in the companion star. In addition, the companion
star is supposed to live as a normal star (e.g., no SN or BH) for a constant 
lifetime $\tau$, no matter what its mass is.  For simplicity, we set the initial 
M$_{\rm BH,0}$ to 40 $M_\odot$ if the Pop III star particle mass 
(M$_{\star}$) is more massive than 40 $M_\odot$, or 10
$M_\odot$ if $10 M_\odot < M_{\star} < 40 M_\odot$.  We do not
consider XRB formation in Pop III stars smaller than 10
M$_\odot$.  The initial BH mass should be a free parameter in the model. 
However, considering that it introduces too much complication into the accretion 
and X-ray spectrum modeling, we simply fix them here. The BH then 
accretes matter from the companion star at the Eddington limit during 
the lifetime of the companion star $\tau$, which we take to be a free parameter in our model, 
or until all the mass of the companion star accretes onto the BH. 
We will discuss the effects of different $\tau$ on the X-ray luminosity 
in the next section. We do not consider the accretion from 
environment, so once the BH ceases accretion from the 
secondary star, its X-ray luminosity is zero.  When electron scattering dominates 
opacity, the isotropic luminosity from accretion is limited to the Eddington luminosity,
\begin{equation}
  L_{\rm Edd} = 1.3 \times 10^{38} \left(\frac{M_{\rm
        BH}}{M_\odot}\right) \unit{erg s}^{-1}
\end{equation}
The radiation efficiency of accretion is $\epsilon$ = $L/\dot{M}c^2$, so
the mass accretion rate at the Eddington limit is $\dot{M}_{\rm Edd}$ =
$L_{\rm Edd}/\epsilon c^2$. Then, the mass evolution of the BH is

\begin{equation}
M_{\rm BH} = M_{\rm BH,0}  \exp\left(\frac{t}{t_{\rm Edd}}\right)
\end{equation}
where the Eddington time $t_{\rm Edd} = M_{\rm BH}/\dot{M}_{\rm Edd} =
\epsilon c^2 M_{\odot} / 1.3 \times 10^{38} \unit{erg s}^{-1} \sim 440
\epsilon \unit{Myr}$ before the companion star runs out of matter.

Since we only update the X-ray luminosity of snapshots between large 
time steps ($\delta z~\sim~0.5$), which is $\sim$ 6 and 12 $\unit{Myr}$ at
$z=20$ and 15, respectively, we use the time averaged luminosity as 
the luminosity of each Pop III binary.
The total BH accreted mass is
\begin{equation}
  M_{\rm acc} = \left\{ \begin{array}{r@{\quad}l}
       M_{\rm BH,max}  - M_{\rm BH,0}  & \textrm{for}\; M_{\star} > M_{\rm BH,max} \\
      M_{\star} - M_{\rm BH,0} & \textrm{for}\; M_{\star} \leq M_{\rm BH,max}
  \end{array}
\right. 
\end{equation}  
where $M_{\rm BH,max}$ = $M_{\rm BH,0} \exp\left(\frac{\tau}{t_{\rm Edd}}\right) $
is the maximum mass of a BH after accreting at the Eddington limit for the
lifetime $\tau$.

The total radiation energy from the accretion process is $\epsilon
M_{\rm acc}c^2$, so the average luminosity of each binary is simply
$\epsilon M_{\rm acc}c^2$/$t_{\rm acc}$, where the accretion time
$t_{\rm acc}$ is $\tau$ for $M_{\star} > M_{\rm BH,max} $, or 
$\ln(M_{\star}/M_{\rm BH,0}) t_{\rm Edd}$ for $M_{\star} \le M_{\rm BH,max} $. 
There is no delay between BH and Pop III formation considered; 
each Pop III binary luminous X-ray radiation is at this constant rate since the Pop III 
star forms in the simulation for its accretion time $t_{\rm acc}$. Using this model, the 
X-ray energy output is simply determined by the three free parameters, 
lifetime $\tau$, binary probability $\alpha$, and radiation efficiency $\epsilon$.

The propagation of X-ray radiation and the effects on the IGM are
dependent on the photon energy because the \hi~cross-section
approximately decreases rapidly as $\sim$ $\nu^{-3}$.  Although, in post-processing, 
we consider a monochromatic spectrum and not a spectral energy distribution 
(SED), we can estimate the effects of different XRB SEDs by
exploring different photon energies. We adapt a multi-color disk (MCD)
blackbody \citep{Mitsuda84} plus a high-energy power-law model to
model the radiation spectrum from Pop III binaries with BHs with
masses of tens of \Ms.  This model was also adopted for mini-quasars
in a cosmological context in \citet{Kuhlen05}, who considered the
photoheating and photoionization effects from a 150 \Ms~BH that has
a softer SED than the binaries presented in this work.  Their model
included a MCD component with luminosities equally divided between a
multi-color disk and a power law component with a power index $\alpha$
both with the same low-energy cutoff. In a MCD model,
each annulus of a thin accretion disk radiates as a blackbody with a
radius-dependent temperature, $T(r) \propto r^{-3/4}$, and the
temperature of the innermost portion of the disk decreases slowly with
BH mass \citep{Makishima00} as $T_{\rm in} \sim 1.2 (M_{\rm BH} / 10
M_\odot)^{-1/4} \unit{keV}$.
The inner disk temperature is about 1.2 keV and 0.8 keV for a 10
\Ms~and 40 M$_\odot$ BH, respectively.  We use a similar SED functional
form as \citet{Alvarez09}, assuming L$_\nu \propto \nu$ for $h\nu < 400
\unit{eV}$, $L_{\nu} \propto \nu^{-1}$ for $400 \unit{eV} < h\nu < 10
\unit{keV}$, and L$_\nu$ = 0 for $h\nu > 10 \unit{keV}$, which has a
mean photon energy of $\sim$770 eV. To study the effects of different
photon energies, we consider several different monochromatic photon
energies ($E_{\rm ph}$) between 300 eV to 3 keV, covering most of the
emissions from this model.


\section{X-ray Luminosity from Rarepeak}
\label{Xrayluminosity}

We present the evolution of the X-ray luminosity from Pop III
sources and its distribution among halos in our Rarepeak survey
volume in this section. We first need to choose the free parameters 
for our X-ray model. We try two companion star lifetimes $\tau = 10, 30 \unit{Myr}$.
We set the probability for a Pop III star becoming an XRB to $\alpha$ = 0.5. 
This assumes that there is a high binary fraction, comparing to 0.36 in \citet{Stacy13},
and almost all Pop III binaries are XRBs.  The radiation
efficiency depends on the BH properties. It is 0.057 for a Schwarzschild BH, 
and increases to $\sim$ 0.4 for a prograde disk around a maximally 
rotating BH \citep{Thorne74}, and could be even higher for a 
magnetized disk \citep{Gammie99}. Here, we simply set it to $\epsilon$ = 0.2. 
We choose this number higher that other studies \citep[e.g.,][]{Ricotti04,Alvarez09}, 
making  the growth of BH slower and the total X-ray radiation 
weaker ($\sim$ 20\% comparing to $\alpha$ = 0.1).
Since both the binary lifetime $\tau$ and XRB probability only impact the X-ray 
luminosity linearly, our results are easily adjusted with different parameters 
when better constraints are available. 

\begin{figure}
\begin{center}
\epsfig{file=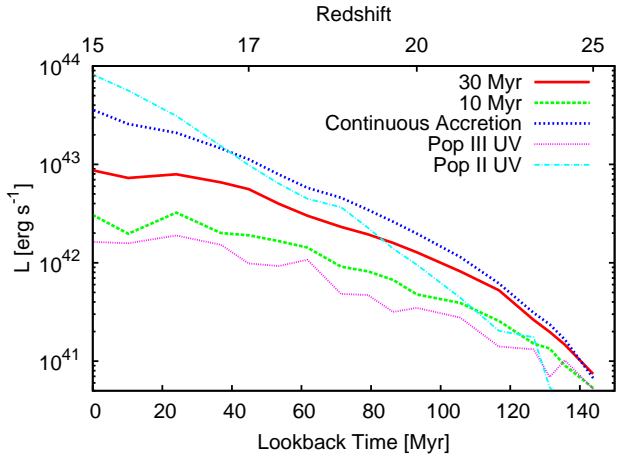,width=1.0\columnwidth}
\end{center}
\caption{Evolution of total X-ray luminosities from Pop III binaries in 
the Rarepeak region with two assumed lifetimes of the binary companion stars. 
In the case of all Pop III BH continuously accreting at the Eddington limit (the blue dotted curve),
their total luminosity is about four times that of the 30 Myr case at $z=15$.    
The luminosities of hydrogen ionizing photons from Pop II and Pop III stars in
the simulation are shown for reference. The UV radiation from Pop II stars dominates
the radiation feedback from $z \sim 19$.
\label{fig:X-ray_evolution}}
\end{figure}

Figure \ref{fig:X-ray_evolution} shows the evolution of the
total X-ray luminosity for two different binary lifetimes $\tau = 10,
30 \unit{Myr}$. We also plot the UV hydrogen ionizing photon
luminosities from both Pop III and Pop II stars to compare with
X-rays.  The X-ray luminosity
follows the Pop III formation rate \citep[see Figure 1 in][]{Xu13}
and thus the UV luminosity of Pop III stars. The total X-ray output
steadily increases until the simulation ends at $z=15$, where $L_{\rm X}
= 8 \times 10^{42} \unit{erg s}^{-1}$, which is consistent with that in
\citet{Mirabel11}.  The X-ray dependence on the binary lifetime is
simple and linear, so that the luminosity from the 30 $\unit{Myr}$ case is
always 2--3 times higher than the $\tau = 10 \unit{Myr}$ scenario.
X-ray radiation from Pop III binaries dominates the total luminosity
budget until the most massive halos begin to form Pop II stars
efficiently.

X-ray production from Pop III stars is debatable because of the
uncertainties in models of metal-free binary evolution.  To calculate
the upper limit of the X-ray luminosity in our simulation, we consider
the unlikely case where BHs in Pop III binaries continuously accrete
at the Eddington limit after formation. This is shown as a dotted blue line 
in Figure \ref{fig:X-ray_evolution}.  At $z > 20$,
this optimistic scenario is comparable to the $\tau = 30 \unit{Myr}$
case, because at that redshift, most Pop III binaries are still
active. At $z=15$, it is about four times higher than the $\tau = 30
\unit{Myr}$ model, suggesting that we are not underestimating the
X-ray output significantly even in the worst scenario. 

\begin{figure}
\begin{center}
\epsfig{file=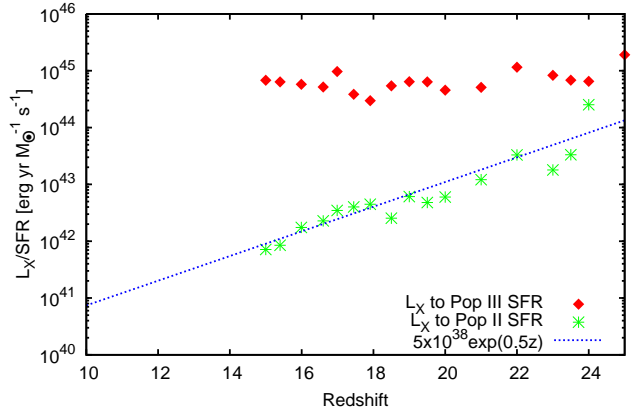,width=1.0\columnwidth}
\end{center}
\caption{Ratios of X-ray luminosity to the Pop III and Pop II star formation rates.
\label{fig:X-ray_SFR}}
\end{figure}

Next, we compute the relationship between the X-ray luminosity and Pop
II and III star formation rates (SFRs), which are shown in Figure
\ref{fig:X-ray_SFR}, for the $\tau = 30 \unit{Myr}$ case.  X-ray
output closely follows the Pop III star formation as $L_{\rm X} \sim 5
\times 10^{44} \unit{erg s}^{-1}$ (SFR$_{\rm III}$/M$_\odot$
yr$^{-1}$).  This is expected in our model because X-ray luminosity is
proportional to the Pop III SFR within the past 30 $\unit{Myr}$.  The X-ray
luminosity can also be correlated to the Pop II SFR over time, $L_{\rm X}
\sim 5 \times 10^{38} \exp(0.5z) (\textrm{SFR}_{\rm II}/ \Ms
\unit{yr}^{-1}) \unit{erg s}^{-1}$ in the range $z = 15-25$, which decreases 
in time simply due to the increasing Pop II SFR.  By
extrapolating this relation to lower redshifts, we can determine that
$L_{\rm X}$ from Pop III binaries become comparable to the scaling
relation found in local starburst galaxies \citep[e.g.,][]{Oh01}, which
happens at $z \sim 9$ when $L_{\rm X} \sim 5 \times 10^{40}
(\textrm{SFR}_{\rm II}/ \Ms \unit{yr}^{-1}) \unit{erg s}^{-1}$.

A correction factor $f_{\rm X}$ for X-ray efficiency is commonly used
in the literature to relate the SFR and the X-ray
luminosity. Using the same definition as \citet{Furlanetto06},
\begin{equation}
  L_{\rm X} = 3.4 \times 10^{40} f_{\rm X}
  \left(\frac{\textrm{SFR}}{M_{\odot} \unit{yr}^{-1}}\right) \unit{erg s}^{-1},
\end{equation}
we find that $f_{\rm X} \simeq 1.5 \times 10^4$ and $f_{\rm X} \simeq
0.015 \exp(0.5z)$ for Pop III and Pop II stars, respectively. While $f_{\rm X}$ for Pop II stars 
is lower than that for Pop III stars (due to Pop II-dominated SFRs), it is still much larger than $f_{\rm X}$(Pop II) 
for normal galaxies. Low-redshift starbursts were estimated to have $f_{\rm X}$ $<$ 1.7 \citep[e.g.,][] {Oh01,Fragos13}, 
while here, $f_{\rm X}$(Pop II) = [27-330] at z = [15-20].

\begin{figure*}
\begin{center}
\epsfig{file=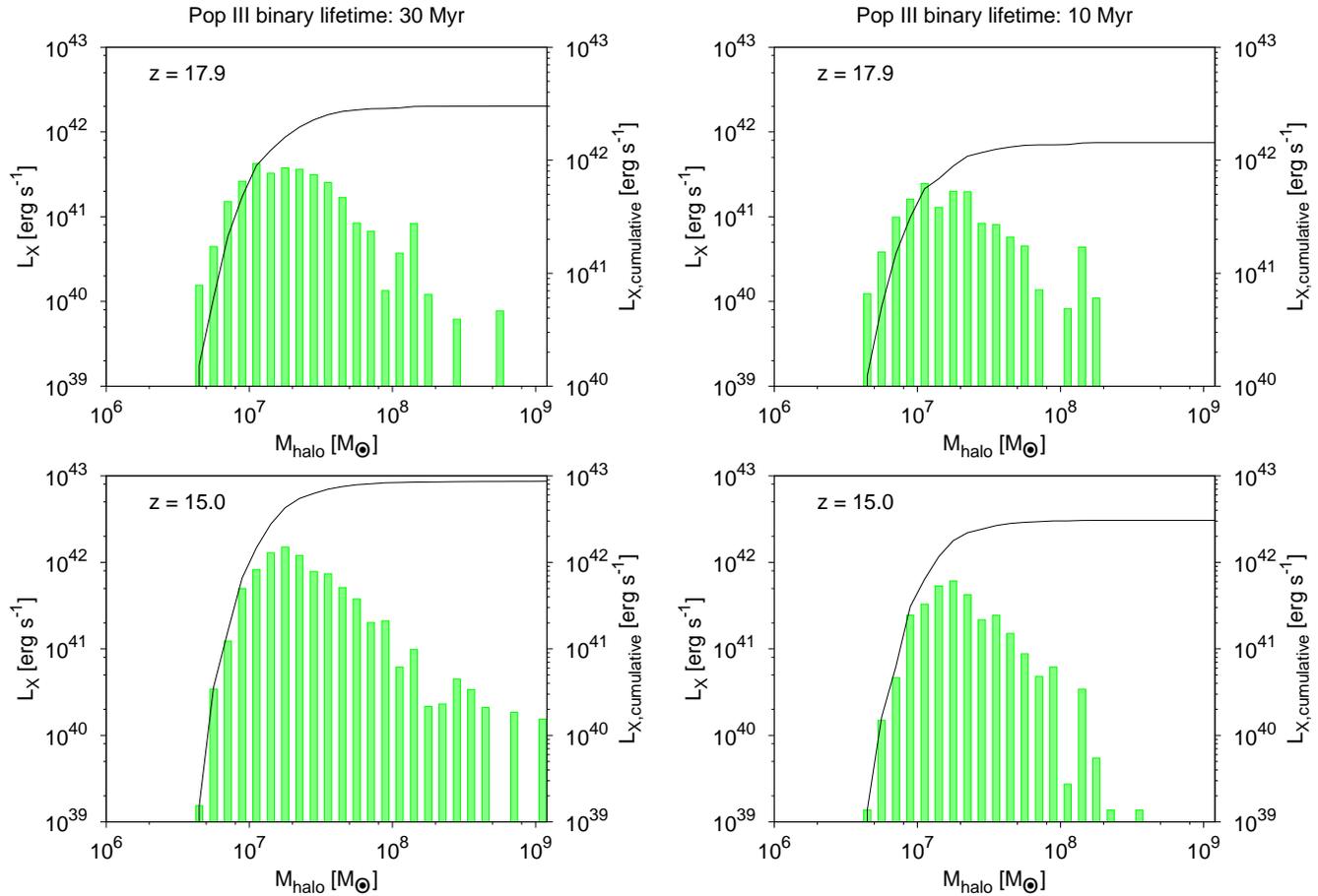,width=1.0\textwidth}
\end{center}
\caption{Distributions of X-ray luminosity from Pop III binaries over halo mass in 
the Rarepeak region of 10 Myr (right) and 30 Myr (left)  cases at z=17.9 (top) and z=15 (bottom). 
The black line is the cumulative X-ray luminosity.
\label{fig:X-ray_histogram}}
\end{figure*}

The distributions of X-ray luminosity from Pop III binaries among
halos at $z = 15$ and $17.9$ for the $\tau = 10$ and $30 \unit{Myr}$
models are shown as functions of halo mass in Figure
\ref{fig:X-ray_histogram}.  Unsurprisingly, the distributions reflect
the same halo mass dependence as the Pop III stars and remnants 
(see Figure 3 in \citet{Xu13}). The
peak of X-ray luminosity is at $\sim$ 2 $\times$ 10$^7$ M$_\odot$ and
most of the X-rays are from small halos that have a small neutral
column density, allowing for a large X-ray escape fraction.  

We now turn our attention to the details of the resulting X-ray radiation
field, using \moray~, adaptive ray tracing \citep{Wise11}, to calculate 
the X-ray propagation into the IGM.


\section{X-ray Heating, Ionization, and Escape Fractions}
\label{raytracing}

Here we present results from calculating the X-ray radiation field by
ray tracing its propagation to better understand how it photoheats
and photoionizes the host halos, the Rarepeak region, and the IGM of
the entire simulated volume. The long mean free paths of X-rays make it
currently computationally unfeasible to trace rays from all the
sources self-consistently within the original time-dependent 
cosmological simulation.  To work around this
limitation, we introduce some approximations and post-process our
simulation to calculate the X-ray radiation field and to study its
effects on the thermal and ionization state of the gas.

Because UV radiative transfer was included in the simulation, we only
consider the transport of X-rays in post-processing.  Starting with
each output of the full simulation, we consider the matter field to be
static and equal to the output at time $t^n$ and allow the chemistry
and energy solvers to evolve the ionization and thermal states of the
gas, using the calculated X-ray radiation field, until the time
$t^{n+1}$ of the next simulation output.  Furthermore, we approximate
multiple X-ray sources within a given halo as a single point source at the halo center with a
luminosity representing the sum of all active X-ray Pop III binaries in
each halo.  For the X-ray luminosity of a halo in each post-processing
timestep, we calculate the average luminosity of the halo as $\langle
L_{\rm X, halo} \rangle \equiv E_{\rm X} / (t_1 - t_0)$, where $E_{\rm
  X}$ is the total X-ray energy by all sources in the halo that is
emitted between the earliest formation time $t_{0}$ and latest
cessation time $t_{1}$ of X-ray-emitting Pop III binaries.%
\footnote{The earliest formation and latest cessation times are
  restricted by the time elapsed in the post-processing timestep,
  i.e., $t_0 = \max(t_0, t^n)$ and $t_1 = \min(t_1, t^{n+1})$.} %
We should mention that this approach results in a slightly different (a few percent)
total luminosity as in the previous section, but
this difference does not manifest into appreciable changes in the
thermal and ionization states of the gas, especially when considering
the uncertainty in X-ray luminosities of Pop III binaries.  Finally,
we consider absorption by neutral and singly ionized helium, which is
important at these high energies, which was neglected in the stellar
UV radiation transport in our original calculation.

\begin{figure*}
\begin{center}
\epsfig{file=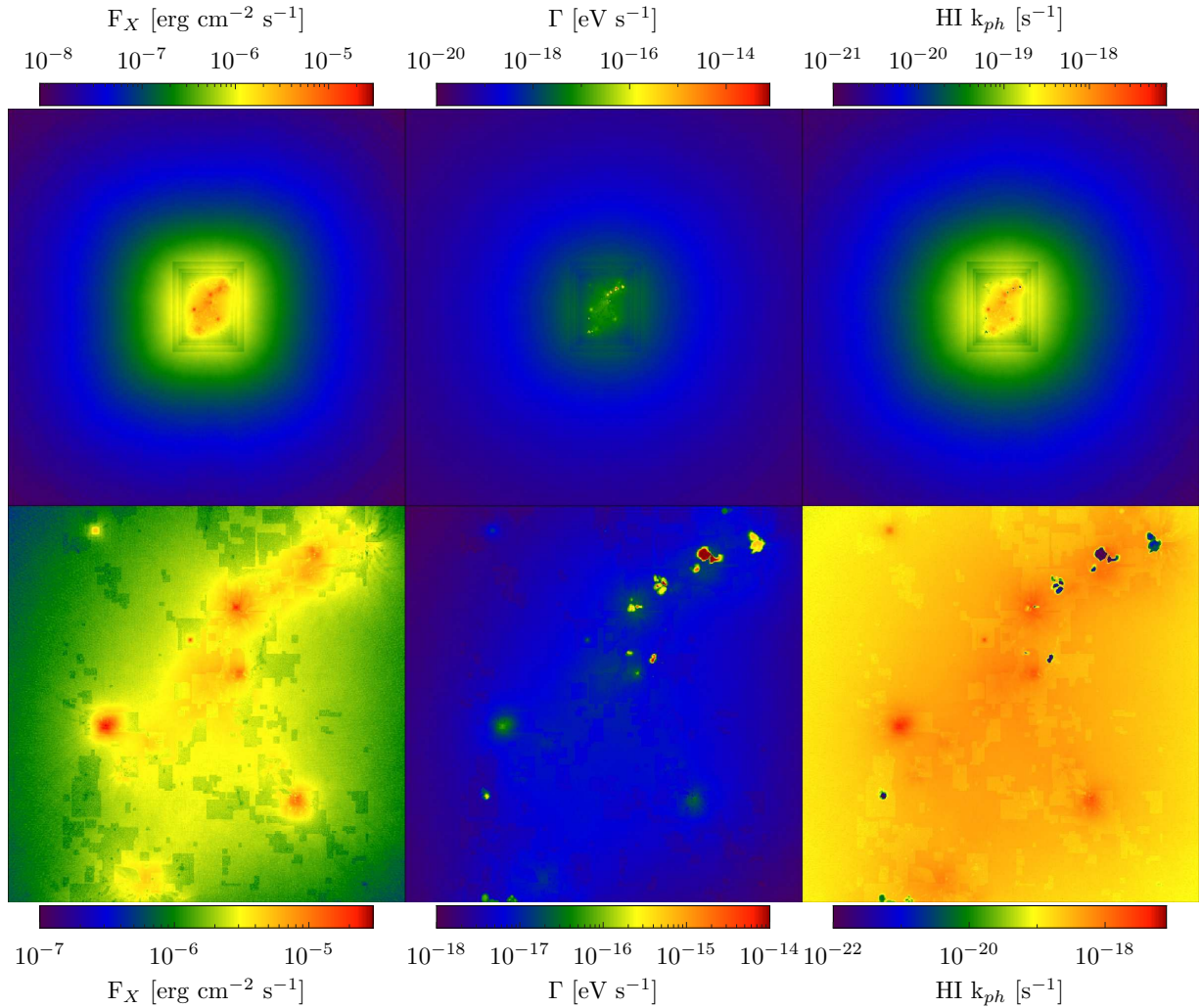,width=0.9\textwidth}
\end{center}
\caption{ Slices of X-ray flux $F_X$ (left), X-ray photoheating rate $\Gamma$ (middle) and X-ray
\hi~photoionization rate k$_{ph}$ (right) of the ray-tracing run with 1 keV photon energy at $z=15$. 
The top panels are images through the entire box  (40 comoving Mpc on a side), and the bottom panels 
show the zoomed-in central squares of 5 comoving Mpc on a side. Out of the high-density region, Rarapeak 
works just like an X-ray point source. The zoomed-in images show some \hii~regions with high
photoheating rates and low \hi~ionization rates. Because the high-resolution cells are not
sampled by enough rays after ray splitting turned off at a large radius, there 
are some grid artifacts in thin layers at the refined and nested region boundaries, as well as in the
zoomed-in images.
\label{fig:X-ray_slice}}
\end{figure*}

We run the ray tracing post-processing using a single energy group of
300 eV, 500 eV, 1 keV or 3 keV, covering most of the spectrum of XRBs 
with tens to one hundred $M_\odot$ BHs.  In Figure \ref{fig:X-ray_slice}, 
we show the slices of X-ray flux, photoheating rate $\Gamma$ and
\hi~photoionization rate $k_{\rm ph}$ through the center of the
simulated volume to illustrate the distribution and effects of the
X-rays for the 1 keV case at $z=15$. There are grid artifacts at the refined
and nested grid boundaries. This is due to the smaller cells which are not sufficiently
sampled by enough rays \citep{Wise11} as we turn off the ray splitting once a ray has
traveled more than 320 comoving kpc to save memory. These artifacts do not affect our
results since they only occur in thin layers. As expected, X-rays can easily
travel out of the hosting halos and Rarepeak region into the normal
IGM. Although X-ray sources inside Rarepeak show complicated
structures, the X-ray radiation field outside of the Rarepeak are
quite spherically symmetric.  The ionization and heating timescales
are sufficiently low so that the IGM properties do not change
considerably, and the X-rays travel through the IGM passively,
eventually reaching a steady state well before the next
post-processing timestep.

\begin{figure}
\begin{center}
\epsfig{file=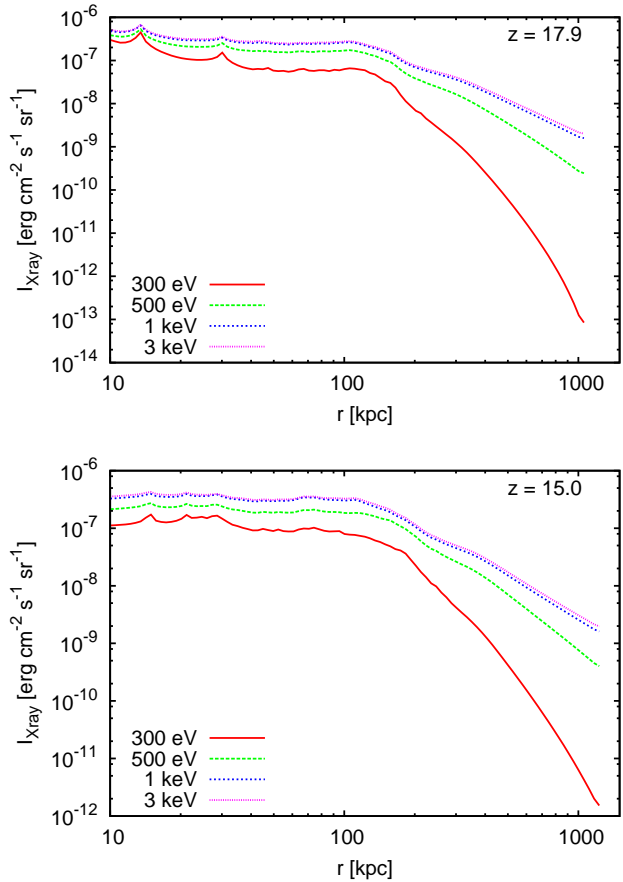,width=1.0\columnwidth}
\end{center}
\caption{Volume-averaged distribution of X-ray intensities as a function 
of the distance to the center of the simulated volume at z=17.9 (top) and z=15 (bottom). 
The X-rays are ray traced by post-processing. Different lines
represent the results from different photon energies, while the luminosity
of the sources are the same.  
\label{fig:X-ray_profile}}
\end{figure}

\begin{figure}
\begin{center}
\epsfig{file=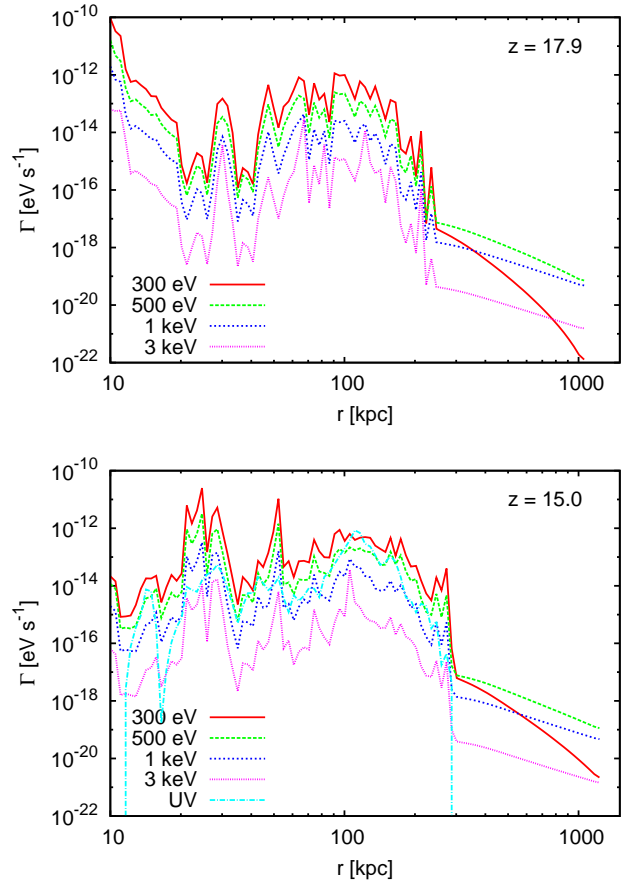,width=1.0\columnwidth}
\end{center}
\caption{Same as Figure \ref{fig:X-ray_profile}, but for the X-ray photoheating rate $\Gamma$.  We also plot
the profile of the UV photoheating rate at z=15 from the original simulation.
\label{fig:Gamma_profile}}
\end{figure}

\begin{figure}
\begin{center}
\epsfig{file=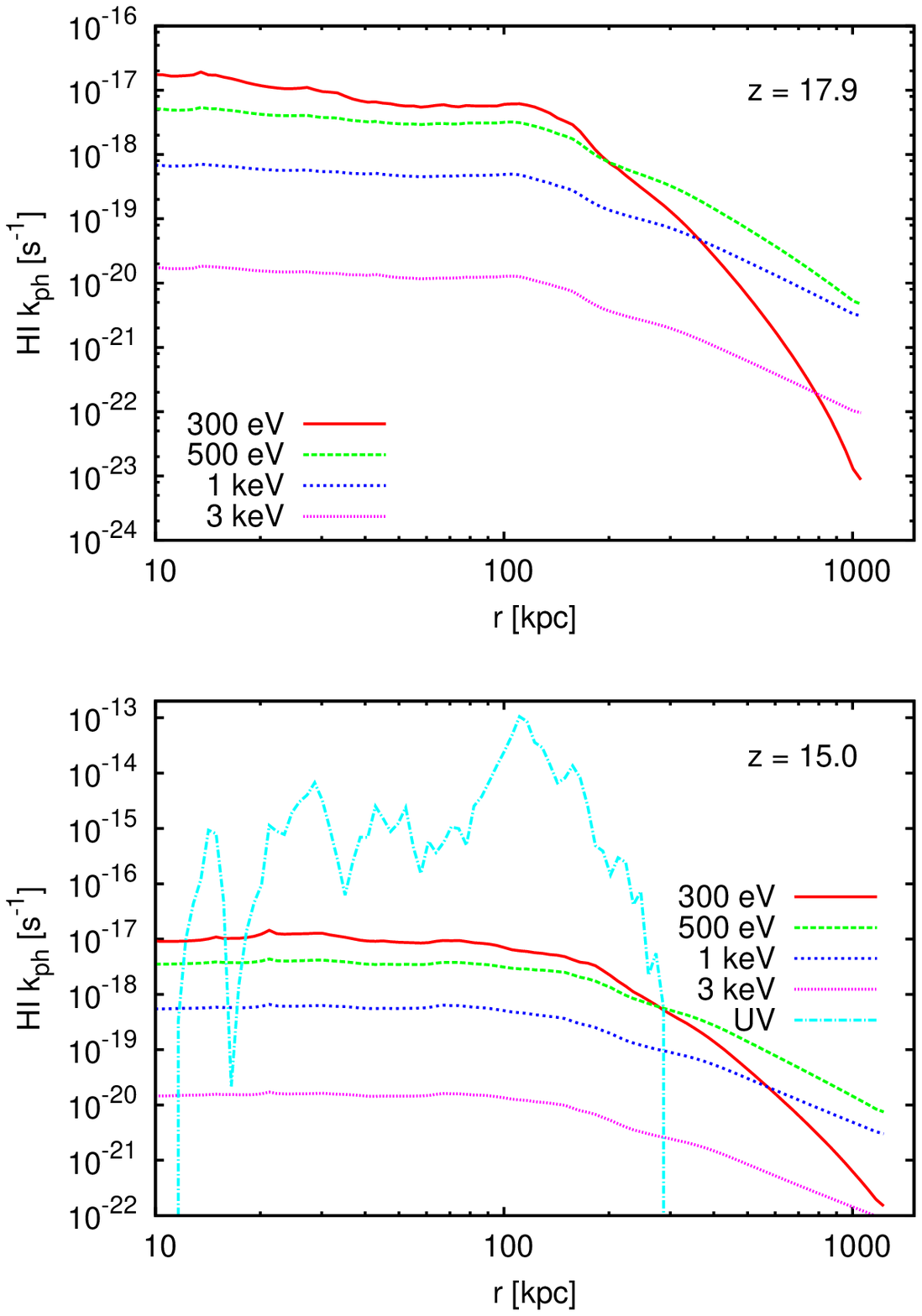,width=1.0\columnwidth}
\end{center}
\caption{Same as Figure \ref{fig:X-ray_profile}, but for the X-ray \hi~photoionization rate k$_{ph}$.
We also plot the profile of the UV photoionization rate at z=15 from the original simulation.
\label{fig:kph_profile}}
\end{figure}

The volume-averaged radial profiles of X-ray intensity,
photoheating rate $\Gamma$ and \hi~photoionization rate $k_{\rm ph}$
for different photon energies are shown in Figures
\ref{fig:X-ray_profile}, \ref{fig:Gamma_profile},
\ref{fig:kph_profile}, respectively. No obvious evolution is observed for 
these quantities from the two shown redshifts, $z=17.9$ and 15. There are very 
different features between the results for hundreds eV ("sub-keV") and 
keV photons. 

\subsection{X-Ray Intensity}

X-ray intensities are almost flat inside the Rarepeak and decrease externally 
because we do not consider star formation outside of the Rarepeak.  

1. Sub-keV Photons. The 300 and 500 eV X-ray intensities are 
much weaker than keV ones because the absorption is
stronger at lower photon energies. Their intensities drop fast outside of the 
source region. Only small amounts of photons escape to the 
IGM. More specifically, at $z=15$, there are about 15.8\% and 50.5\% 
of the total X-ray energy escaping the Rarepeak region at 200 kpc
radius (3.2 comoving Mpc) for 300 eV and 500 eV photons,
respectively. At 1 Mpc (16 comoving Mpc) radius, the escape fractions are
only 0.11\% and 13.1\%, respectively. 

2. KeV photons. The attenuation of the keV photons
through the IGM is weak, resulting in their intensities dropping just
slightly faster than $1/r^2$ geometric dilution.  This leads to a
significant amount of radiation at keV energy scales escaping into the
IGM and possibly contributing to the X-ray background.  
At $z=15$, there are about 73.9\% and
81.0\%, of the total X-ray energy escaping the Rarepeak region at 200 kpc
radius (3.2 comoving Mpc) for 1 keV and 3 keV cases, respectively. 
At 1 Mpc (16 comoving Mpc) radius, the escape fractions are 44.3\% and 53.2\%,
respectively.
 
 \subsection{X-Ray Photoheating}
The radial distribution of X-ray photoheating rates is more
complicated than that of X-ray intensities. There is a large variance 
inside the Rarepeak because the heating rates are highest in 
star-forming halos, which have high electron fractions and thus 
have most of the photon energy deposited to thermal energy.  

1. 300 eV Photons. Because the attenuation of low-energy photons
through \hi, \hei, and \heii~are much higher (recall that their
cross-sections scale as $\sim$ $1/E_{\rm ph}^3$), high heating rates are
created by these low-energy photons, which extend to the boundary of the
Rarepeak region. The X-ray photoheating can significantly change the 
thermal state of the gas inside the Rarepeak in just millions of years.

2. 500 and 1 keV Photons.  Photons at this energy range heat the Rarepeak
region more weakly, but can efficiently heat the IGM when compared to lower-energy 
photons. However, their overall rates are not high, and they need a long time to  
significantly increase the IGM temperature.  

3. 3 keV Photons.  Though the 3 keV photon intensity is strong, weak 
interactions between high-energy photons  and gas result in very low 
heating rates. 

We also plot the photoheating rate profile at $z=15$ from
the UV ionizing photons that were originally included in the
simulation to illustrate the significance of X-ray heating. The
spherically averaged UV heating rate is comparable to the X-ray
photoheating inside the Rarepeak and drops to almost zero out of the
star-forming region because the UV photons are all absorbed within a
few kpc of their origins.  Because the UV radiation can only heat the
gas within these small-scale \hii~regions, the thermal morphology from
UV radiation is porous, whereas the X-ray radiation creates a smoothly
varying component of the heated IGM.
 
\subsection{X-Ray Photoionization}
The distribution of \hi~photoionization rates $k_{\rm ph}$ for
different photon energies is similar to the photoheating rate
$\Gamma$, but their profiles are much smoother inside the Rarepeak.

1. 300 eV Photons. The 300 eV photons have relatively high photoionizing 
rates inside Rarepeak and extend to the edge of the Rarepeak. But at 
rates of $\sim$ 10$^{-17}$ s$^{-1}$, they are insignificant compared 
to the UV photoionization for the star-forming 
region of Rarepeak.

2. 500 and 1 keV Photons. Similar to our photoheating results, radiations 
in the 0.5--1 keV band are more efficient at photoionizing the IGM, 
where the Rarepeak acts as a single source at large distances, still at very 
low rates $<$ 10$^{-18}$ s$^{-1}$.  

3. 3 keV Photons. Their photoionization rates are much lower compared
to lower photon energy cases in both Rarepeak and the IGM.

We also plot the profile of the photoionization rate at
$z=15$ from the UV radiation. Contrary to the behavior in the heating
rates, the photoionization by UV is much more efficient than the
ionization by X-rays inside the Rarepeak, as a higher fraction of UV
photon energies is used to ionize the IGM than to heat them. Outside
of the Rarepeak region, the UV radiation is fully attenuated, and the
photoionization rates drop to zero accordingly.


\section{Effects of X-rays on the IGM}
\label{effects}

Ray tracing of X-rays is very computationally expensive because it is
in the optically thin limit.  Thus, it is not possible to perform the
ray tracing self-consistently with all other physics and chemistry for
the duration of the simulation.  However, in the previous section, we
showed that the X-ray ionization rates are low enough so
that it does not dynamically affect the electron fraction of the IGM and then 
the opacity of X-rays and any ensuing star formation.  In this limit, it is 
safe to assume that the X-ray intensity will remain constant 
in between outputs of the original simulation.  We thus freeze 
the X-ray radiation field and use a simple model to study the 
X-ray photoheating and photoionization of the IGM.
 
\subsection{One Zone Model of X-Ray Heating and Ionization}
\label{onezone}

\begin{figure}
\begin{center}
\epsfig{file=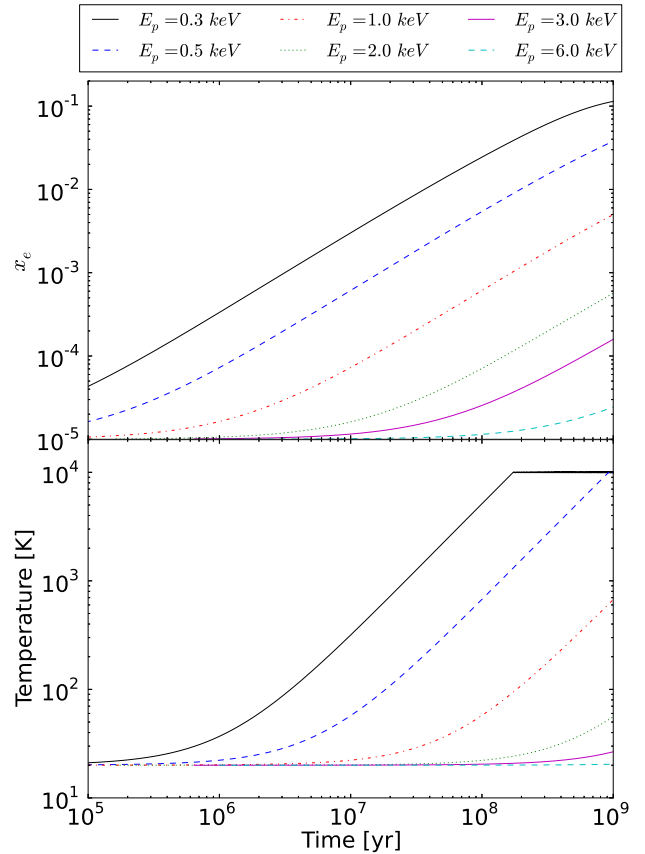,width=1.0\columnwidth}
\end{center}
\caption{Evolution of the electron fraction (top) and temperature 
(bottom) from a one-zone calculation for different photon
energies with the same X-ray flux 10$^{-6}$ erg cm$^{-2}$ s$^{-1}$. 
\label{fig:one_zone_eph}}
\end{figure}

\begin{figure*}
\begin{center}
\centerline{
\mbox{\epsfig{file=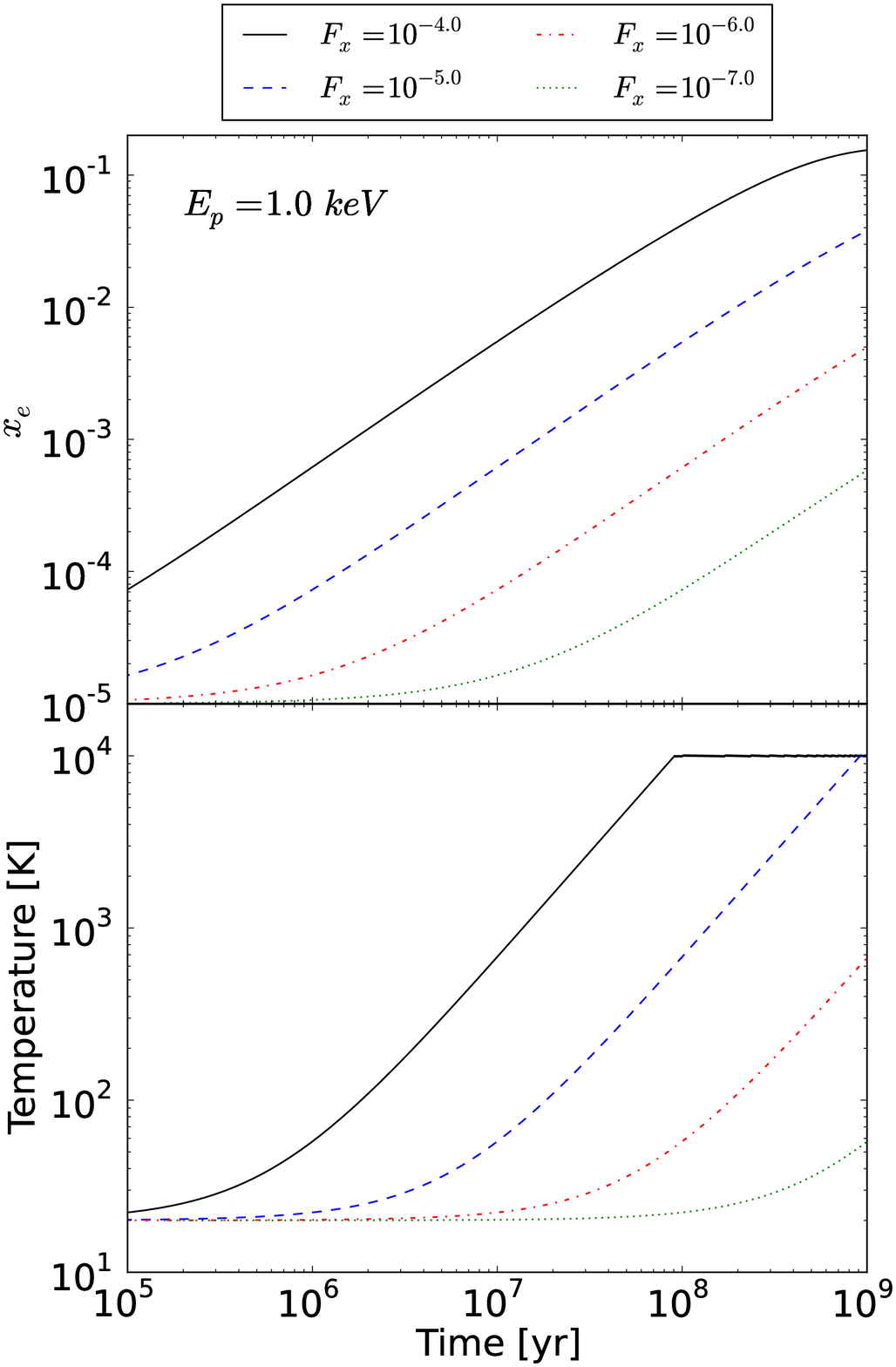,width=0.5\textwidth}}
\mbox{\epsfig{file=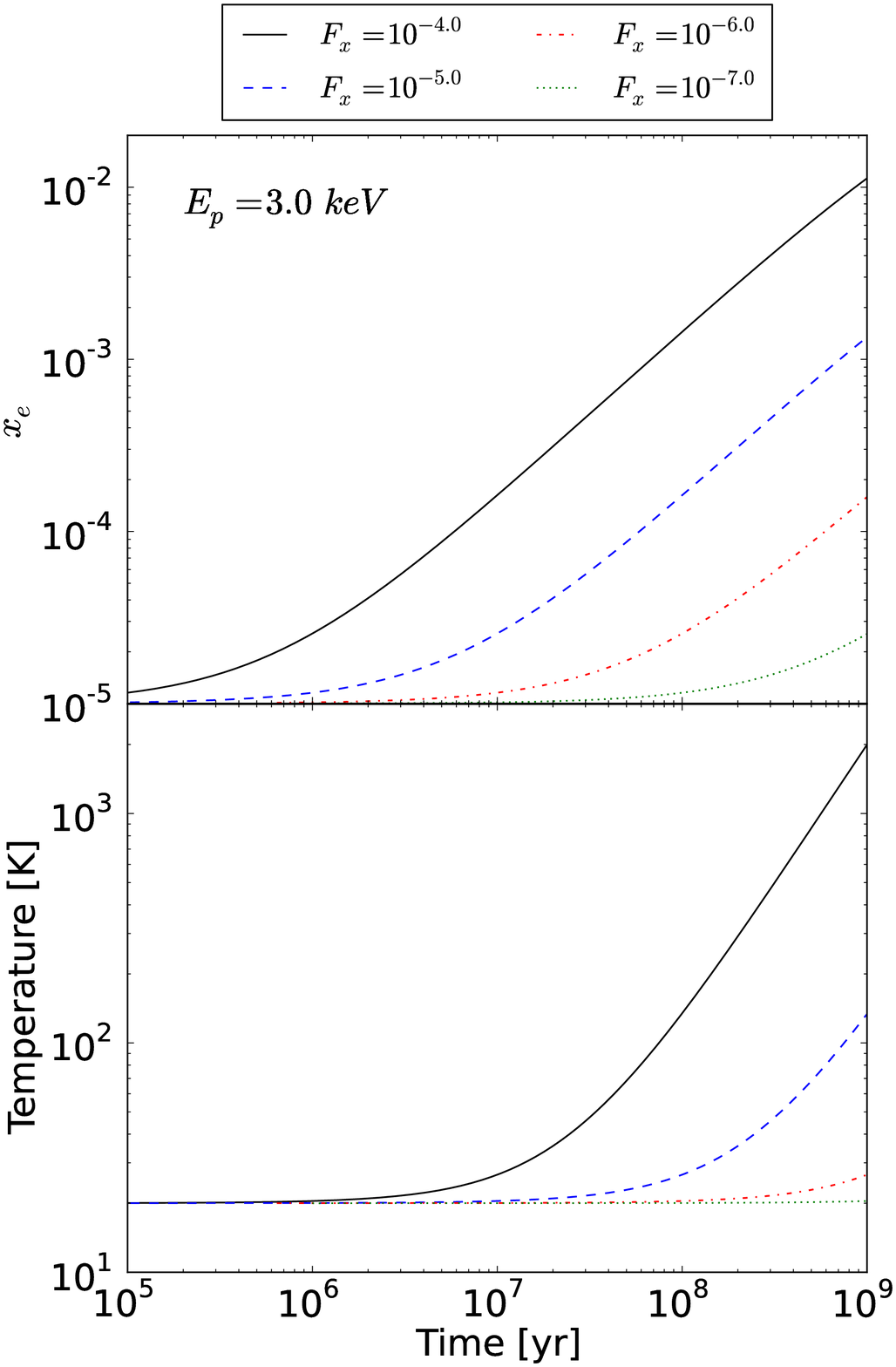,width=0.5\textwidth}}
}
\end{center}
\caption{Evolution of the electron fraction (top) and temperature (bottom) from
one-zone calculations of 1 keV and 3 keV photon energies with a variety of 
radiation fluxes.
\label{fig:one_zone_flux}}
\end{figure*}

We first use a one-zone model to study the effects of X-rays of
different photon energies and fluxes on the IGM. This model includes
X-ray heating and ionization with secondary ionizations, collisional
ionizations, recombinations (case B), and primordial atomic
cooling. The changes of electron fraction (x$_e$) and thermal energy (E$_{th}$) are expressed as
\begin{eqnarray}
\frac{dx_e}{dt} &=& (1-x_e)(k_{ph}+n_e k_1)-x_e n_e \alpha_B \\
\frac{dE_{th}}{dt} &=& \Gamma_{ph} - \Lambda 
\end{eqnarray}
Here, the photoionization and photoheating rates k$_{ph}$ and $\Gamma_{ph}$ with the X-ray
secondary ionizations use the results from \citet{Shull85}, the same as in our ray tracing 
calculation \citep{Wise11}.  k$_1$ is the collisional ionization coefficient in \citet{Abel97}, 
using the fit in \citet{Janev87}. The case B recombination coefficient, $\alpha_{B}$ is 
2.59 $\times$ 10$^{-13}$ (T/10$^4$ K)$^{-0.7}$ cm$^{-3}$ s$^{-1}$  \citep{Osterbrock89}. 
The primordial atomic cooling rate $\Lambda$ is computed using the table in \citet{Sutherland93} 
when the gas temperature is over 10$^4$ K.  

The model considers a constant X-ray flux impacting an initial cold
($T = 20$ K) and neutral ($x_e = 10^{-5}$) gas parcel with mean cosmic density.  
The calculation starts at $z=15$ and evolves for 1 Gyr.  The photon energy
used in the calculation is not redshifted and is kept as a
constant. Also, the decrease of temperature due to the expansion of
the universe is not considered.  

We first study the effects of different photon energies with an X-ray 
flux of 10$^{-6}$ erg cm$^{-2}$ s$^{-1}$, which is close to the X-ray fluxes generated
by Rarepeak in our ray-tracing simulations. 
We plot the evolution for the electron fraction and temperature in 
Figure \ref{fig:one_zone_eph} for this X-ray flux of different photon energies
in a range from 300 eV to 6 keV. 

1. Sub-keV Photons. X-rays below 1 keV can significantly heat and ionize 
the IGM. The 300 eV case heats the gas to 10$^4$ K in $\sim$ 100 Myr, 
at which point atomic hydrogen cooling becomes 
efficient, resulting in an ionization boost of 10 percent, which is 
enough to contribute a non-negligible amount to the optical depth 
from Thomson scattering.

2. KeV Photons. The keV photons can only moderately heat the gas 
and weakly ionize the IGM. For 1 keV photons, even for 1 billion years, 
the gas temperature only reaches slight over 600 K and the electron fraction is just
5 $\times$ 10$^{-3}$, not high enough to impact the optical depth of the CMB 
from Thomson scattering. But this electron fraction is high enough to stimulate the $H_2$ formation, 
and then, in turn, the Pop III star formation \citep{Ricotti01,Yoshida04}. For even higher photon energies, the heating 
and ionizing effects are negligible at this level of X-ray flux.

For keV radiation,  stronger fluxes are needed to heat 
and ionize the IGM to a meaningful level. The evolution of the electron fraction and 
temperature is plotted in Figure \ref{fig:one_zone_flux} for 1 keV and 3 keV 
photon energies with various X-ray fluxes.  X-ray radiation of 1 keV photons 
with a flux of 10$^{-5}$ erg cm$^{-2}$ s$^{-1}$ can barely heat the gas to 10$^4$
K and ionize it to about 4\%. This shows that for increasing the X-ray
flux, the heating is more significant than the associated
ionization. For 3 keV photons, even this elevated X-ray flux is not
sufficient to heat and ionize the gas within 1 Gyr.

\subsection{X-Ray Background Modeling}

In a mean patch of the IGM, a small opacity for X-ray photons of
$E\gtrsim1\,{\rm keV}$ renders the mean free path truly cosmological
($\sim$ a hundred comoving Mpc for 1 keV photons and several comoving 
Gpc for 3 keV photons). Therefore, we need to consider the X-ray
background caused by sources outside the simulation box, in addition
to the local transfer of X-ray photons.  This is supported by our
results in Section \ref{raytracing}, showing that large amounts of keV
X-rays easily escape out of their host halos, the Rarepeak region, and
even the entire simulation box, which then contribute to the X-ray
background. In this subsection, we describe our method that calculates
the X-ray background from Pop III binaries.\footnote{We show in
   Section \ref{raytracing} that when the ``normal'' X-ray luminosity, which
  is calibrated from low-redshift galaxies \citep[e.g.][]{Fragos13},
  is assigned to metal-enriched stars, its contribution to the X-ray
  background is negligible compared to that from Pop III
  binaries. Therefore, we simply ignore the X-rays from metal-enriched
  binaries in our simulation. }

Our simulation only allow stars formed inside our refined Rarepeak
region, which is only 0.22\% of the entire survey volume, containing
0.37\% of the total baryon mass.  To compensate for these unresolved
sources outside the Rarepeak region, we simply use the density
distribution to populate X-ray sources out of the star-forming region
in our simulation. We do not intend to establish a fully
self-consistent correlation between X-ray luminosity from Pop III
binaries and the underlying baryon density in such a coarse
resolution, but our aim is to populate this region with X-ray sources
in the most likely places.  We first project the X-ray luminosity and
baryon density to the root grid of 512$^3$. For the cells with X-ray
sources, we calculate the mean X-ray luminosity $\overline L_{X,0}$
and baryon density $\overline \rho_0$. Then, for each cell in the
region external to the Rarepeak with a baryon density higher than the
mean density $\overline \rho_0$, we assume there is an X-ray source
with luminosity $\overline L_{X,0}$.  We find that sources populated
in this manner produce X-ray luminosities that are several times higher
than in the Rarepeak region alone.  The total X-ray luminosities of the
entire simulated volume are 6.7, 3.1, and 8.5 times those of Rarepeak
region at $z=24$, 17.9, and 15, respectively.

In principle, the background itself can have spatial fluctuation due
to the inhomogeneous distribution of radiation sources outside the
simulation box. Nevertheless, simulating the structure formation on
such a large scale, while resolving star-forming regions with the
relevant local astrophysical processes in detail is almost impossible
in practice at this time.  Because inhomogeneity in the source
distribution at large lookback times will be observed to be almost
uniform inside the box, we simply treat the background sources to be
uniformly distributed. We also assume that the entire simulation box is
a good representation of the average universe and takes its luminosity
as the mean, globally averaged luminosity.

The (proper) X-ray background intensity $J_{\nu}$ (${\rm erg\,
  s^{-1}\, cm^{-2}\, Hz^{-1}\, sr^{-1}}$) at an observed frequency
$\nu$ and redshift $z_{{\rm obs}}$ will then be given by the
following:
\begin{equation}
  J_{\nu}(z_{{\rm obs}})=\left(1+z_{{\rm
        obs}}\right)^{3}\int_{0}^{\infty}\frac{dr_{{\rm
        os}}}{1+z_{{\rm s}}}\,\bar{j}_{\nu_{{\rm s}}}(z_{{\rm
      s}})\,\exp\left[-\tau_{\nu_{{\rm obs}}}\right],
\label{eq:Jnu_back}
\end{equation}
where $\bar{j}_{\nu_{{\rm s}}}$ (${\rm erg\, s^{-1}\, cm^{-3}\,
  Hz^{-1}\, sr^{-1}}$) is the comoving emission coefficient given by
\begin{equation}
  \bar{j}_{\nu_{{\rm s}}}=\frac{1}{4\pi}\frac{L_{\nu_{{\rm s}},{\rm
        Box}}(z_{{\rm s}})}{V_{{\rm Box}}},    
  \label{eq:jnu_com}
\end{equation}
where $L_{\nu_{{\rm s}}}$ (${\rm erg\, s^{-1}\, Hz^{-1}}$) is the proper, total
luminosity inside the simulation box at source frequency $\nu_{{\rm
    s}}$ and redshift $z_{{\rm s}}$, and $V_{{\rm Box}}$ is the
comoving volume of the simulation box. The optical depth originates
from the absorption of photons by HI and HeI:
\begin{eqnarray}
  \tau_{\nu_{{\rm obs}}} & = & \tau_{\nu_{{\rm obs}},{\rm
      HI}}+\tau_{\nu_{{\rm obs}},{\rm HeI}}\nonumber \\
  & = & \int_{t(z_{{\rm s}})}^{t(z_{{\rm obs}})}c\, dt\, \nonumber \\ 
   &    & \times \left\{ \
    n_{{\rm HI}}(z_{{\rm s}})\,\sigma_{{\rm HI}}(\nu_{{\rm s}},\,
    z_{{\rm s}})+n_{{\rm HeI}}(z_{{\rm s}})\,\sigma_{{\rm
        HeI}}(\nu_{{\rm s}},\, z_{{\rm s}})\right\} \nonumber \\
  & = & \int_{z_{{\rm s}}}^{z_{{\rm obs}}}\frac{c\, dz_{{\rm
        s}}\,(1+z_{{\rm s}})^{-5/2}}{H_{0}\sqrt{\Omega_{m}}}  \nonumber \\
   & & \times  \left\{n_{{\rm HI}}(z_{{\rm s}})\,\sigma_{{\rm HI}}(\nu_{{\rm s}},\,
    z_{{\rm s}})+n_{{\rm HeI}}(z_{{\rm s}})\,\sigma_{{\rm
        HeI}}(\nu_{{\rm s}},\, z_{{\rm s}})\right\} ,  \nonumber \\
    & &
  \label{eq:tau}
 \end{eqnarray}
where we neglect absorption by HeII, because the ionization fraction 
of the IGM at $z \ge 15$ remains very low. $r_{{\rm os}}$ is the comoving 
line-of-sight distance that a photon has traveled given by
\begin{equation}
  r_{{\rm os}}=\frac{2c}{H_{0}\sqrt{\Omega_{m}}}\left\{ (1+z_{{\rm
        obs}})^{-1/2}-(1+z_{{\rm s}})^{-1/2}\right\} ,
  \label{eq:ros}
\end{equation}
and the redshifted frequency $\nu/\nu_{{\rm s}}=(1+z_{{\rm
    obs}})/(1+z_{{\rm s}})$.  Over the frequency range of our
interest, $100\,{\rm eV}\lesssim h\nu\lesssim3\,{\rm keV}$, absorption
cross-sections are well approximated by power laws:
\begin{equation}
\sigma_{{\rm HI}}(\nu_{s})=6.5\times10^{-14}\,\left(\frac{h\nu_{s}}{{\rm eV}}\right)^{-3.25}\label{eq:sigma_HI}
\end{equation}
and
\begin{equation}
\sigma_{{\rm HeI}}(\nu_{s})=1.55\times10^{-12}\,\left(\frac{h\nu_{s}}{{\rm eV}}\right)^{-3.22}.\label{eq:sigma_HeI}
\end{equation}
With these cross-sections, Equation (\ref{eq:tau}) becomes
\begin{eqnarray}
 \lefteqn{ \tau_{\nu_{{\rm obs}}} = \left(\frac{\Omega_{b}}{0.044}\right)\left(\frac{h}{0.7}\right)\left(\frac{\Omega_{0}}{0.27}\right)^{-0.5}\left(\frac{1+z_{s}}{1+25}\right)^{1.5} } \nonumber \\
&  & \times \left\{4.10863\left(\frac{X}{0.75}\right)\left(\frac{h\nu_{0}}{{\rm
        keV}}\right)^{-3.25}\left[\left(\frac{1+z_{s}}{1+z_{{\rm
            obs}}}\right)^{1.75}-1\right]\right.\nonumber\\
 & & +  \left. 10.13573\left(\frac{Y}{0.25}\right)\left(\frac{h\nu_{0}}{{\rm
       keV}}\right)^{-3.22}\left[\left(\frac{1+z_{s}}{1+z_{{\rm
           obs}}}\right)^{1.72}-1\right]\right\}, \nonumber \\
 & &          
 \label{eq:tau_simple}
\end{eqnarray}
where $X$ and $Y$ are mass fractions of hydrogen and helium, respectively.

As we approximate the X-ray SED with a monochromatic frequency
$\nu_{0}$ and the bolometric X-ray luminosity $L_{0}$
($\equiv\int_{\rm X-ray}L_{\nu_{{\rm s}}}d\nu_{{\rm s}}$), Equation
(\ref{eq:Jnu_back}) can be simplified as:
\begin{eqnarray}
  L_{\nu_{{\rm s}},{\rm Box}}(z_{{\rm s}}) & = & L_{0,{\rm
      Box}}(z_{{\rm s}})\,\delta^{D}(\nu_{{\rm s}}-\nu_{0})\nonumber\\
  & = & L_{0,{\rm Box}}(z_{{\rm s}})\,\frac{1+z_{{\rm
        obs}}}{\nu}\delta^{D} \left(1+z_{{\rm s}}-\frac{1+z_{{\rm
          obs}}}{\nu}\nu_{0}\right), \nonumber \\
   & &        
  \label{eq:Lnu_delta}
\end{eqnarray}
and thus using Equations (\ref{eq:ros}) and
(\ref{eq:Lnu_delta}),
\begin{equation} 
  J_{\nu}(z_{{\rm
      obs}})=\left(\frac{\nu}{\nu_{0}}\right)^{3/2}\frac{c\left(1+z_{{\rm
          obs}}\right)^{3/2}}{4\pi \nu_{0}
    H_{0}\sqrt{\Omega_{m}}}\,\frac{L_{0,{\rm Box}}(z_{{\rm
        s}})}{V_{{\rm Box}}}\,\exp\left(-\tau_{\nu_{{\rm
          obs}}}\right),
  \label{eq:Jnu_delta}
\end{equation}
where $1+z_{{\rm s}}= (\nu_{{\rm 0}}/\nu)(1+z_{{\rm obs}})$ is
implied.

In practice, Equation (\ref{eq:Jnu_delta}) should be integrated in piecewise 
frequency intervals due to the time-discrete nature of the simulation. 
Especially for the contribution from the most recent past to the 
present because spatial fluctuation will be non-negligible, 
we calculate the fluctuating 3D X-ray background by adopting 
a scheme by \citet{Ahn09}. We locate the 3D field of X-ray luminosities 
frozen at the most recent past in the full box periodically and calculate 
the contribution at every location of the box by summing over the full 
contribution from all the X-ray sources but within the corresponding 
lookback time. Since this includes an out-of-box contribution, it will differ from 
the X-ray intensity distribution calculated from sources only inside the
box at the observed redshift. 

\begin{figure*}
\begin{center}
\centerline{
\mbox{\epsfig{file=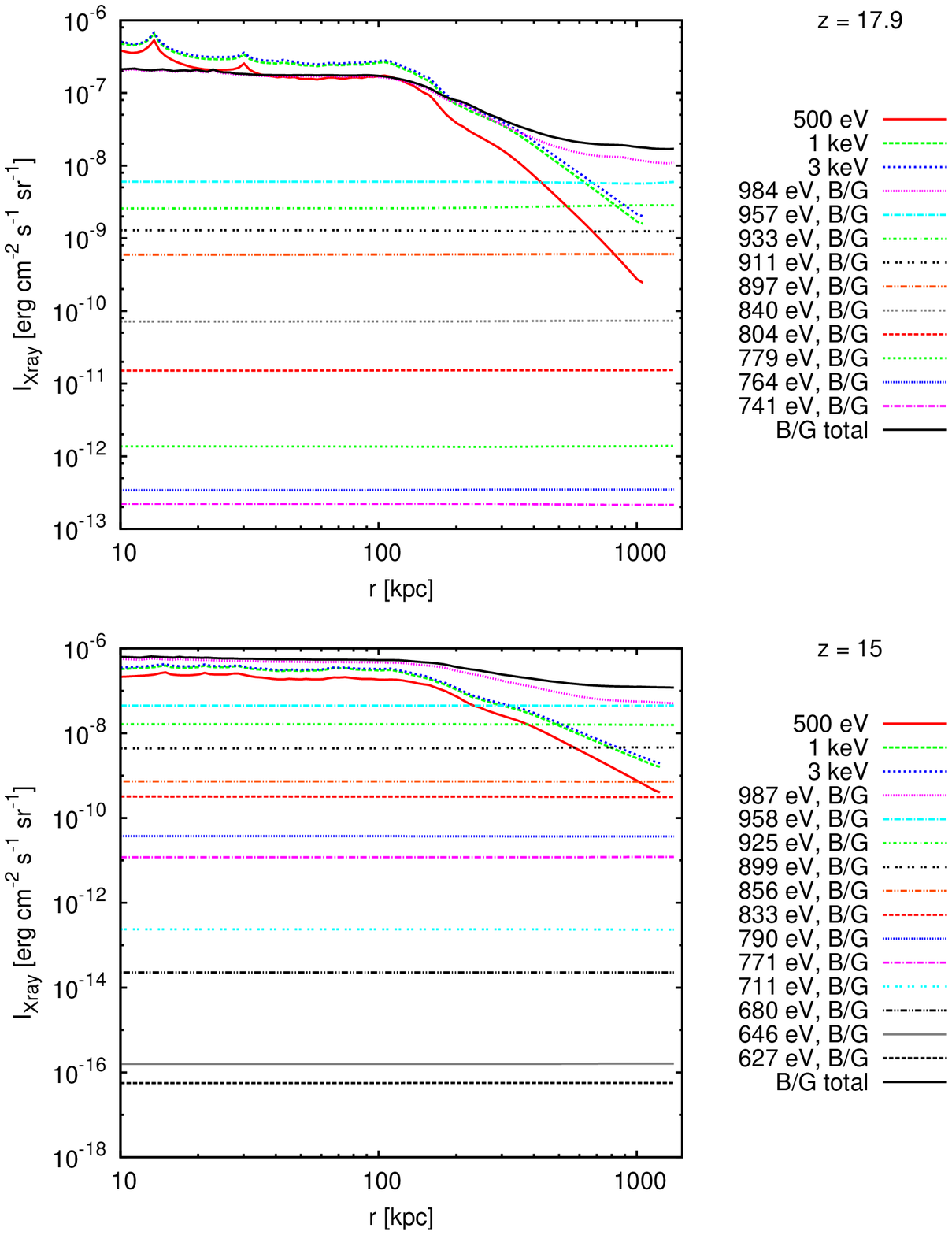,width=0.5\textwidth}}
\mbox{\epsfig{file=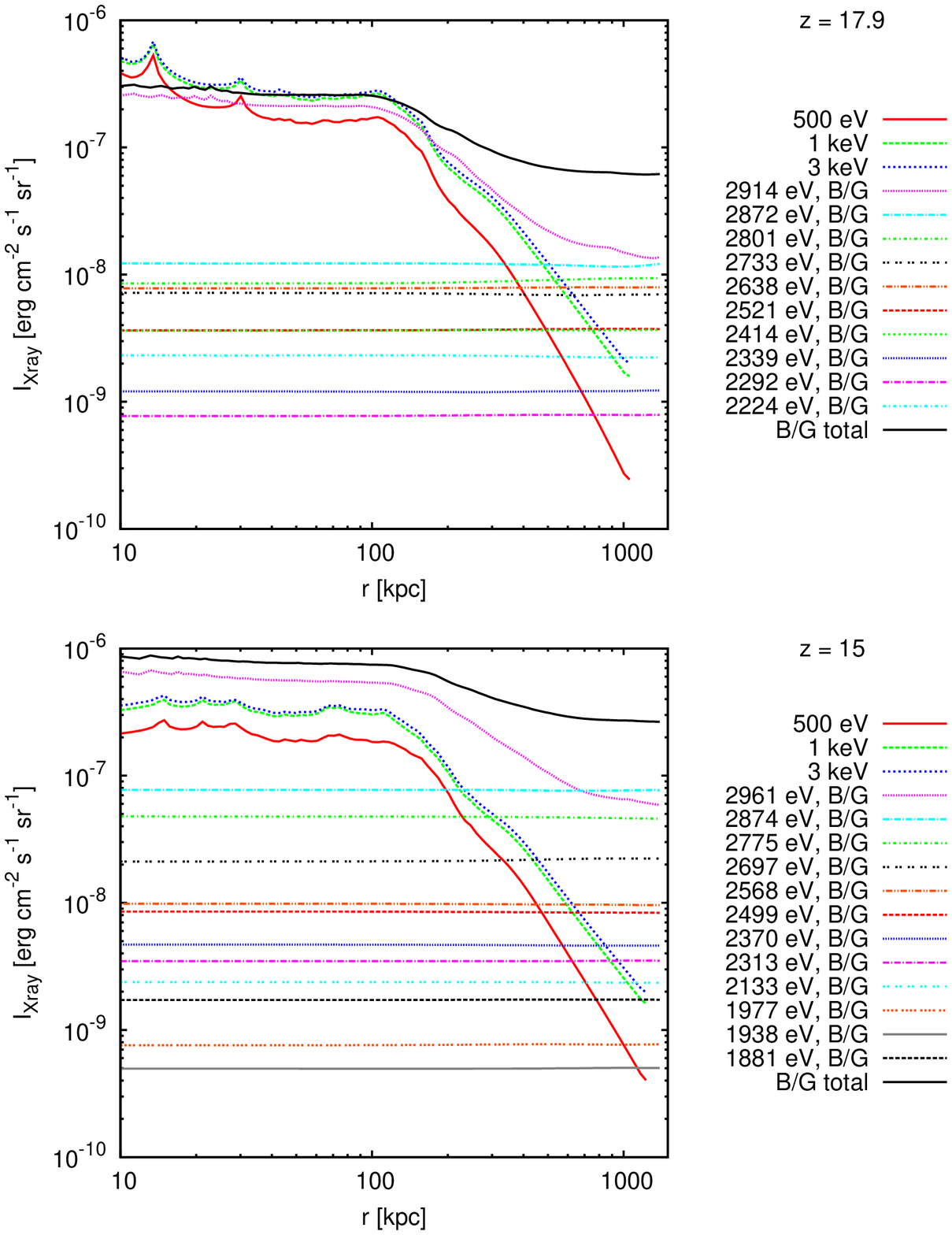,width=0.5\textwidth}}
}
\end{center}
\caption{Distribution of X-ray intensities as function of the distance to the simulated
volume center of 1 (left) and 3 (right) keV photons. The X-ray background from 
earlier redshifts is represented by the redshifted photon energy. For $z=15$, background 
from some redshifts is omitted to make the figures simple. The X-ray intensities 
from ray tracing are plotted to show the relative importance between local sources 
and the background. 
\label{fig:X-ray_background}}
\end{figure*}

\begin{figure}
\begin{center}
\epsfig{file=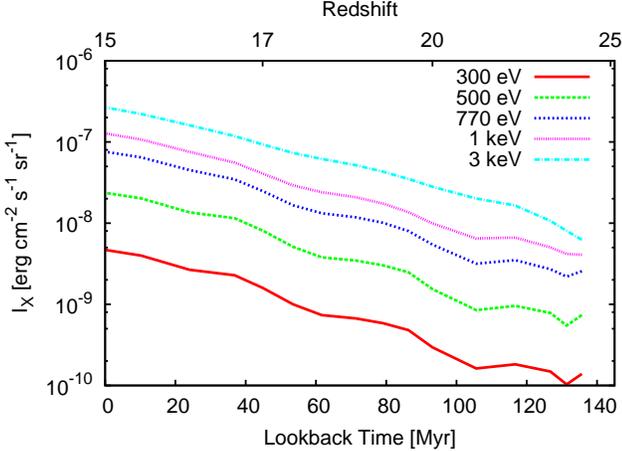,width=1.0\columnwidth}
\end{center}
\caption{Evolution of the volume-weighted averaged intensity of the X-ray 
background of the entire simulated volume. The background from 3 keV photons 
is more than 2 times that from the 1 keV photons. The sub-keV X-rays in the 
simulation box are dominated by the Rarepeak sources and locally distributed, 
so their volumed-weighted intensities are much weaker than those of keV cases. 
\label{fig:XB_evolution}}
\end{figure}

We plot the X-ray background intensities of monochromatic 1 keV and 3
keV X-ray photons as functions of distance to the simulation box
center in Figure \ref{fig:X-ray_background}. The X-ray background
intensities from higher-redshift sources are represented by their
redshifted photon energies, which are all uniform in space except 
for the most recent contribution. We also plot the local X-ray intensities
from our ray-tracing calculation for comparison. 
We also perform the calculations for the sub-keV
X-rays. They show that the X-ray intensities from outside the
simulation volume are negligible because most of their photons are
absorbed locally.

It is interesting to understand the relative importance of the
X-ray effects between the local sources and the X-ray background in
different regions.  Inside the Rarepeak, the X-ray intensities from
local sources are strong and are at the same level as the X-ray
background from nearby sources, whereas outside the Rarepeak, the
X-ray background dominates.  The background outside has an equivalent X-ray
intensity at the same order of magnitude as the Rarepeak region, which has a very high
Pop III stellar density. The X-ray background intensities at $\sim$ 20
comoving Mpc away from the source region are about 10 and 50 times
higher than those from the Rarepeak region alone at redshifts $z=17.9$
and 15, respectively.  Thus, for the X-ray feedback, the mean IGM
should be mainly heated and ionized by the external X-ray background rather
than by any local sources. In addition, as the external background 
resides in lower frequencies, its impact on heating and ionization 
of the IGM is boosted from the sheer value of X-ray intensity by a factor 
of  $\sim$  $(\nu/\nu_0)^{-3}$ .

The X-ray background from 1 keV photons and 3 keV photons is indeed
quite different due to its differing mean free paths.  For the 1 keV
photons, the background is dominated by cosmological nearby sources
because the X-ray radiation from sources at a distance of $dz > 2$ is
negligible. This is consistent with the fact that the mean free paths
for the 1 keV X-ray photons are $\sim$ a hundred comoving Mpc (more
specifically $\sim$ 4 $\times$ 10$^{4}$ /(1+z)$^{2}$ comoving Mpc). Therefore,
the background from the 1 keV case is not building up from these early
redshifts.  At even higher energies, the mean free path for 3 keV photons
is a few Gpc in the mean IGM, resulting in a non-negligible fraction of the
radiation from $z \la 25$ propagating to $z = 15$. However, Since the total X-ray radiation
energy density is increasing during this period, the contributions
from distant ($\Delta z \gg 1$) sources are still much weaker than those from the
nearby sources. 

We show the evolution of averaged intensities of the total X-ray
background from all cases in Figure \ref{fig:XB_evolution}, including
an additional monochromatic X-ray of 770 eV, which is the mean photon
energy from the assumed SED in Section \ref{sec:xray_model}.  At $z=15$,
while the total X-ray luminosity from all sources in the simulation box is 
7.1 $\times$ 10$^{43}$ erg s$^{-1}$, the mean X-ray intensity 
is ranging from 5 $\times$ 10$^{-9}$ to 3 $\times$ 10$^{-7}$ erg 
cm$^{-2}$ s$^{-1}$ sr$^{-1}$ for low to high energy photons.   
The background of 3 keV photon case is more than twice as high as the 1
keV case because of the large mean free path of the higher-energy
photons, and the ratio of background at 3 keV and 1 keV shows little
evolution from $z=25$ to $z=15$.  The average intensities
for the sub-keV cases of the entire simulated box are much weaker
because most of the radiation is confined inside the Rarepeak. The 300
eV background is about two orders of magnitude smaller than that of
the 3 keV case. However, since their attenuation in the IGM is much stronger
(as shown in Figure \ref{fig:one_zone_eph}), their heating and
ionizing effects may still be as important as the higher-energy
background.

At even lower redshifts, X-rays from Pop III binaries
likely grow slowly or even start to drop due to Pop III star formation
being suppressed by metal enrichment \footnote{Though the Pop III star formation rate at 
lower redshifts is still very uncertain, e.g.,  \citet{Crosby13a} suggests that Pop III star formation 
continues on at a pretty steady rate until at least $z=10$.}; the X-ray background of higher-energy 
photons, which survive from higher redshifts, should be more important at later times. 
However, since the heating and ionizing effects are much weaker for the high-energy 
photons as shown in  the previous subsection, the effects of their 
background are still expected to be weak.  We will study the long-term effects
of X-ray background in  the next subsection.

\subsection{Effects of X-Ray Background on the IGM}

\begin{figure*}
\begin{center}
\centerline{
\mbox{\epsfig{file=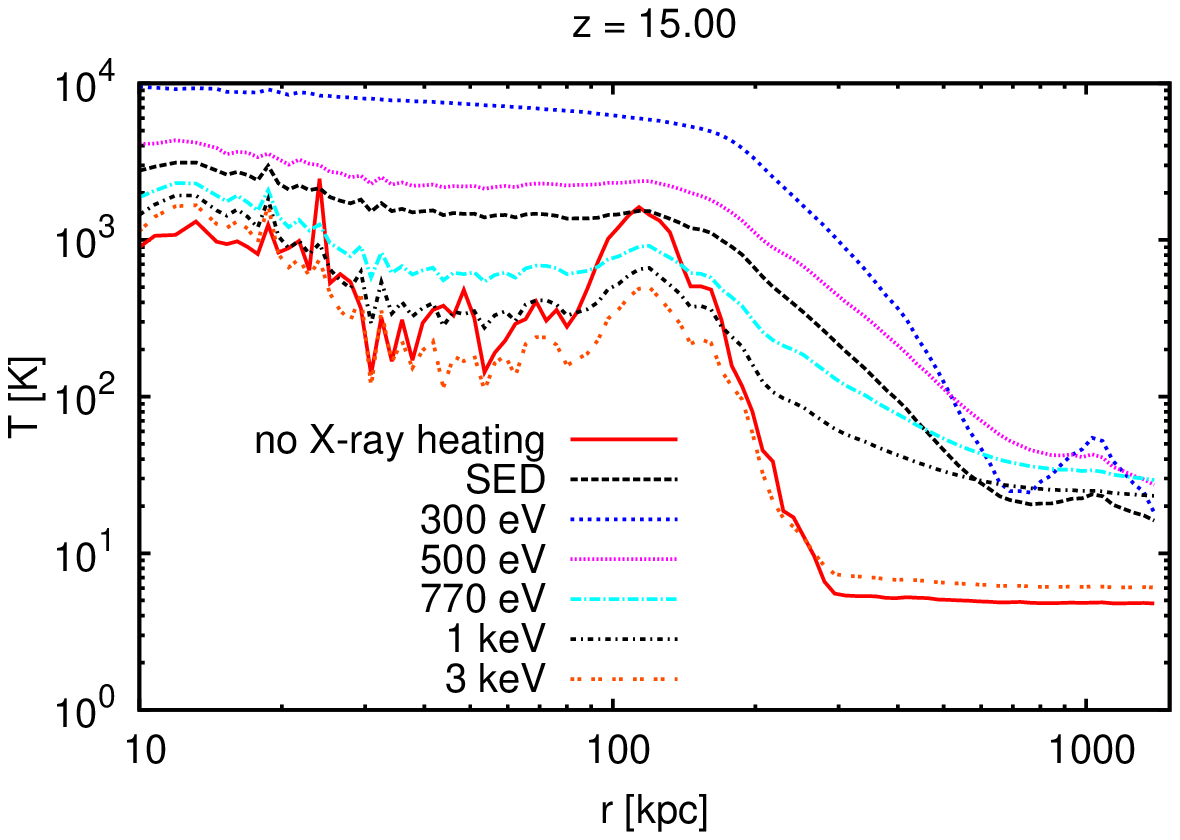,width=0.5\textwidth}}
\mbox{\epsfig{file=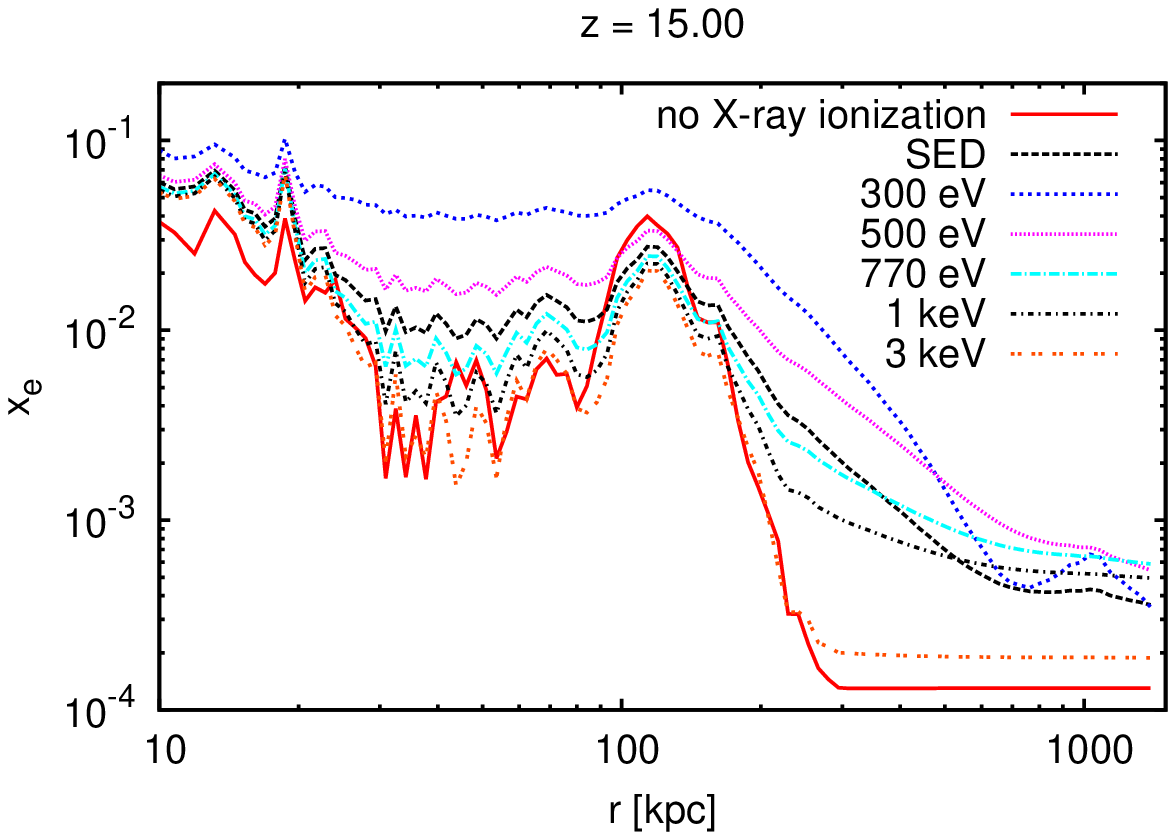,width=0.5\textwidth}}
}
\end{center}
\caption{Volume-weighted averaged radial profiles at z=15 of temperature (left) and electron fraction (right) of different X-ray
photon energies. 
\label{fig:z15_profiles}}
\end{figure*}

\begin{figure}
\begin{center}
\epsfig{file=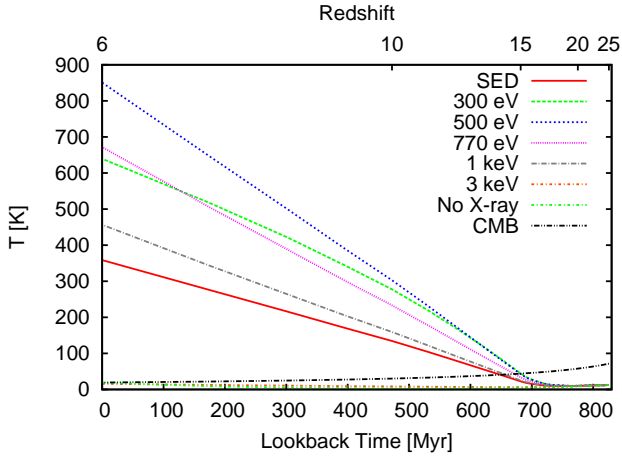,width=1.0\columnwidth}
\end{center}
\caption{Evolution of the averaged IGM temperature to show the heating effect
of the X-rays. We also plot the un-X-ray-heated  temperature and the CMB
temperature for reference. 
\label{fig:Temperature_evolution}}
\end{figure}

\begin{figure}
\begin{center}
\epsfig{file=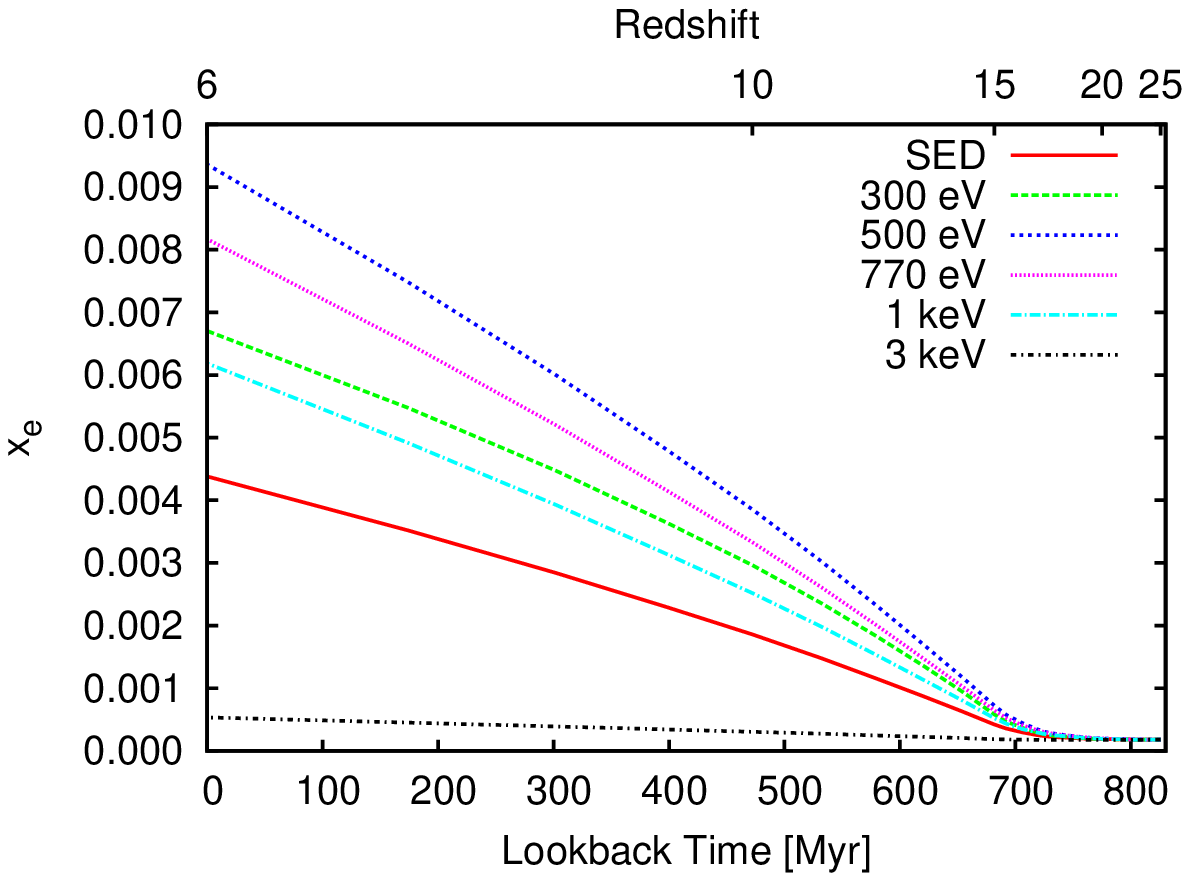,width=1.0\columnwidth}
\end{center}
\caption{Evolution of the averaged electron fraction to show the ionization effect of the X-rays.
\label{fig:ef_evolution}}
\end{figure}

We now estimate the heating and ionizing effects of our X-ray
background model by applying our one-zone model to each 
cell in the 512$^3$ base grid from the full simulation, starting at $z=24$.  At each
snapshot that is separated by $\Delta z \sim 0.5$, we fix the density
distribution and allow the electron fraction $x_e$ and temperature $T$
to evolve according to the one-zone model and the X-ray background
intensity at the given redshift $z_i$ and frequency. The X-ray 
heating and ionization by the redshifted photons with different energies 
and intensities are calculated and summed at each timestep. We include
the adiabatic expansion effect in calculating the temperature change. We then adjust
the temperature and electron fraction in the next snapshot at redshifts
$z_i + \Delta z$ by $\Delta T$ and $\Delta x_e$, respectively. This process 
is repeated until we reach the final redshift $z = 15$ of the simulation.  After
this point, we fix the X-ray background and density distribution and
continue the one-zone calculation for each cell to $z=6$. We 
consider five cases with different photon energies of 300 eV, 500 eV,
770 eV, 1 keV, and 3 keV.  As previously mentioned, the sub-keV 
cases locally heat and ionize the gas instead of adding to
the background intensity. Although all of these calculations 
consider a monochromatic X-ray spectrum, nevertheless, we can obtain 
more realistic synthetic results by averaging the effects weighed by the 
luminosities based on the assumed SED in Section \ref{sec:xray_model}.  

We first show the temperature and electron fraction volume-weighted
averaged radial profiles at $z=15$ in Figure \ref{fig:z15_profiles}.
We also plot the profiles from the original simulation with UV and
other additional heating and cooling processes for comparison.  
Because the UV photoionization and photoheating are not considered in the one-zone model,
the temperatures and ionization fractions at some radii inside Rarepeak are lower than 
the values from the original full simulation.

There are very different heating and ionization patterns for different photon energies.  

\begin{itemize}
\item X-ray photoheating

1. Sub-keV Photons.   The low-energy X-rays significantly heat both inside and outside 
the Rarepeak. Inside the Rarepeak, the X-ray heating pattern is much smoother than 
the individual \hii~regions created by UV photoheating. The heating is so strong that 
the temperatures of the 300 eV case are close to 10$^4$ K, which is limited by atomic hydrogen 
line cooling. Outside, the heating drops with a radius to close to the temperature of the 1 keV case at $r > 1
\unit{Mpc}$. The sub-keV cases are effected by outside (Rarepeak) sources significantly, then 
their radial profiles show bumps at a large radius.  

2. KeV Photons. The heating effect of keV photons is very weak, both inside and outside
the Rarepeak. There is no significant increase in temperature for 3 keV photons, while
the heating of the 1 keV case is slightly more effective, resulting in $\Delta T \sim 20 \unit{K}$, close
to those of sub-keV cases for a normal IGM at a large radius.

3. SED. The synthetic heating is similar to the 500 eV case, but weaker, inside and near 
Rarepeak, then it drops faster to even weaker than that of the 1 keV case at large distances 
($\sim$ 500 proper kpc or 8 comoving Mpc).

\item X-ray photoionization 

The distribution of electron fraction is similar to the distribution of temperature. The electron 
fractions in all cases at $\sim$ 1 Mpc away from the center are smaller than 10$^{-3}$. These ionization
levels are not high enough to contribute to the optical depth to electron Thomson scattering 
of the CMB significantly ($\delta \tau \, \ll \, 0.01$).

1. Sub-keV Photons. The highest electron fractions from the 300 eV case are just below 0.1 
inside the Rarepeak. For the IGM outside of Rarepeak,  the electron fraction of the 300 eV case 
is the highest at $r \la 500 \unit{kpc}$, and then the 500 eV case has the highest
electron fraction. The effect of the 770 eV photons is closer to the 1 keV case than 
to the 500 eV case.

2. KeV Photons.  The ionization effect of keV photons is very weak ($\delta x_e$ $<$ 10$^{-3}$), both inside and outside
the Rarepeak. The photoionization of 1 keV photons decreases slowly with radius due to
their high escape fraction and then close to those of sub-keV cases at $r > 1\unit{Mpc}$.

3. SED. The synthetic ionization is close to the 770 eV case, inside and near the Rarepeak 
(inside $\sim$ 500 proper kpc or 8 comoving Mpc radius), and then is even weaker than 
those of the 1 keV cases at large distances.
 
\end{itemize}

Figures \ref{fig:Temperature_evolution} and \ref{fig:ef_evolution}
show the evolution of volume-weighted averaged temperature and
electron fraction over the entire comoving (40 Mpc)$^3$ simulated volume 
with results from the one-zone calculation of each cell, respectively.  
We exclude the central (8 comoving Mpc)$^3$ region to
obtain the mean values for a typical patch of the universe. For the
temperature evolution, we also show the CMB temperature and the
temperature from the hydrodynamical simulation.  Star formation and
feedback are not considered in the simulation after $z=15$ for
computational reasons.
 
The results of this volumetric average are consistent with our
one-zone calculation that uses a fixed initial conditions and X-ray flux, described
in Section \ref{onezone}.  This confirms that, at the same luminosity, the photon energy  
is the most important determinant in the heating and
ionization history of the IGM. 

\begin{itemize}

\item IGM temperature

Before $z \sim 16$, the X-ray background is still weak, and the heating of the 
IGM is not enough to compensate the decrease of temperature due to the adiabatic
expansion. After that, the X-ray background is strong enough to increase 
the mean IGM temperature, which (except for the 3 keV case) passes the CMB temperature 
at redshifts just below 15, passes 100 K by $z=10$, and then gradually increases 
to a range of  400 to 900 $\unit{K}$ at $z=6$. Some regions, especially for the 
300 eV case, are heated to over 10$^4$ K and then, are cooled by atomic hydrogen, 
resulting in a slowing down temperature growth.

1. Sub-keV Photons. Low-energy X-rays rapidly heat the IGM. The 300 eV
photons significantly affect the regions near the sources, and the temperature
growth slows approaching to $\sim$ 640 K because of a combination of the weak heating
of the IGM outside of the Rarepeak and efficient radiative cooling
near the Rarepeak.  On the other hand, 500 and 770 eV X-rays continue to
heat all the IGM of the simulated volume more effectively to higher than 
850 and 670 K, respectively.

2. KeV Photons.  The heating for the 1 keV  radiation is much more pronounced than that of
the 3 keV case, as its temperature reaches 450 $\unit{K}$. For the 3 keV photons, though their
intensity is more than twice stronger, the heating effect is so weak
that the IGM temperature only increases by a few Kelvin by $z = 6$ 
when compared with the case without X-ray heating.  

3. SED. The heating effect from the synthetic full spectrum is similar to 
that for the 1 keV case at large radii, and, accordingly, the mean 
temperature of the simulation volume is close to that of the 1 keV case.  
The mean IGM is heated to $T \sim 360 \unit{K}$ at $z=6$.

\item IGM Electron Fraction

The evolution of electron fraction is simpler than the evolution of
temperature because the equilibrium between photoionization
and case B recombination is not yet reached in most of the volume
for all cases. 

1. Sub-keV Photons. The 500 eV X-rays are the most efficient in ionizing the 
entire volume because they can propagate farther 
away from their sources than lower-energy photons while having stronger 
interactions with the IGM than higher-energy photons.  Our estimations show 
that, at $z=6$, the maximal change in the volume-averaged electron 
fraction is $\sim$ 0.01.  The electron fractions 
are  smaller for 300 and 770 eV photons, at 0.8\% and just $<$ 0.7\%, respectively. 
Though unlikely, if all of the X-ray radiation exists at $\sim$ 500 eV, its ionizations 
might contribute to the optical depth to the Thomson scattering. 

2. KeV Photons.  The X-ray background from 3 keV photons
is inefficient in ionizing the IGM, only resulting
in $\Delta x_e \sim 10^{-3}$ by $z=6$.  The 1 keV X-rays photoionize
the IGM moderately, to an electron fraction of $x_e \sim 5 \times 10^{-4}$ at $z \sim
15$ and $\sim 6  \times 10^{-3}$ at $z=6$.

3. SED. The mean IGM electron fraction from the synthetic full spectrum is
only $x_e$ $\sim$ $2 \times 10^{-3}$ at $z=10$ and $x_e < 5 \times 10^{-3}$ 
at $z=6$, suggesting that ionizations from an X-ray background are 
a minor correction when calculating the optical depth due to Thomson 
scattering.

\end{itemize}
  
In summary, the IGM can be heated and ionized by the X-ray background from Pop III binaries. 
The photoionization effect is likely unimportant in direct observations, at least to the optical depth to 
Thomson scattering of the CMB, as the electron fraction might not be elevated over 0.005, 
However, this electron fraction could have some really substantial positive 
effects on $H_2$ formation, and cause a general uptick in Pop III SFR 
elsewhere. On the other hand, the temperature of a large volume of the IGM may exceed 100 K
by redshift $z=10$. These heated regions of the IGM might be barely detectable by 21 cm 
SKA observations. The details of the mock 21 cm observations of Rarepeak and the surrounding
area will be presented in a forthcoming paper \citep{Ahn13}.


\section{Discussion and Conclusions}
\label{summary}

In this paper, we utilize a self-consistent cosmological radiation
hydrodynamic simulation of the formation of the first galaxies to
study the luminosity, propagation and effects of X-rays generated by
Pop III binaries. Using the distribution and evolution of more than
13,000 Pop III stars and remnants in 3,000 halos in a survey volume of $\sim$ 138
comoving (Mpc)$^3$ from $z=30$ to $z=15$ and with a simplified Pop III
XRB model, we estimate that there is an X-ray luminosity of
$\sim$ 10$^{43}$ erg s$^{-1}$ from the Rarepeak at $z=15$, equating to
a luminosity density of $\sim 5 \times 10^{40}$ erg s$^{-1}$ per
comoving Mpc$^{3}$. We find that X-rays from Pop III binaries are much
stronger than their UV ionizing radiation at pre-ionizing the
universe, which dominate the photon budget in the early universe
before galaxy formation becomes prevalent.

We study the impacts of these high-energy photons on the IGM in our
simulated box of comoving (40 Mpc)$^3$.  Due to the long mean free
paths of X-ray photons (especially for keV photons) and the
weak heating and photoionizing effects, radiation sources from
cosmological distances and long integration times are needed for an
accurate calculation.  We combine three different methods---ray
tracing, a one-zone model, and X-ray background modeling---for such a
study in order to investigate the X-ray propagation, intensity
distribution, and long-term effects on the IGM temperature and
ionization.  We first post-process the simulation with ray tracing to
study the X-ray distribution through our simulated volume from the
sources inside Rarepeak.  Keeping the luminosity unchanged, we trace
X-rays of 4 different monochromatic photon energies of 300 eV, 500 eV,
1 keV, and 3 keV. While sub-keV X-rays are significantly
absorbed, higher-energy photons easily escape from halos and the high-density 
Rarapeak region.  Thus, we conclude that keV
radiation sources are likely the major contributor to the X-ray
background.  Our work also shows that local sources of X-rays do not
have a significant impact on the typical IGM in a short period of time.  
Even the sub-keV X-rays, which are mostly absorbed locally, do not 
heat and ionize the gas fast and significantly with our calculated X-ray flux,
suggesting that the ionization and thermal state of the IGM at high
redshift is weakly dependent on the X-ray source population.

We estimate the evolution of the X-ray background using the X-ray intensity from our
simulation by assuming some X-ray distribution outside the AMR refined region in our simulated
box and using the mean IGM optical depth for hydrogen and helium.  As
suggested by our ray-tracing approach, only the keV
photons can escape the galaxies and contribute meaningfully to the IGM
X-ray background. For 1 keV photons, their mean free path is on the
order of a hundred comoving Mpc, and only the photons from within
$\Delta z \sim 2$ contribute to the background. For higher-energy
photons, their mean free path is on the order of 1 comoving Gpc, and the
3 keV background includes photons from nearly all radiation sources
$z > 15$, resulting in the 3 keV X-ray background being more than
double that of the 1 keV case.

We apply a one-zone model combining our X-ray background with the IGM
properties from our simulation to estimate the heating and ionizing
effects to the IGM everywhere inside our (40 comoving Mpc)$^3$ box by 
the X-rays before reionization. As expected,
they are very sensitive to the photon energies.  Sub-keV X-rays, which 
only impact the nearby IGM and negligibly contribute to
the background, can significantly heat and ionize the Rarepeak region
and also have moderate heating and ionization effects on the nearby
IGM. They heat the Rarepeak region significantly (300 eV case even to
10$^4$ K), and may have important impacts on the ongoing star and
galaxy formation.  X-rays of $\sim$ 1 keV, which can escape the
galaxies and Rarepeak region, also have moderate effects to the heat
and ionize the IGM, and may contribute substantially to $H_2$ formation 
in distant regions.  The cool IGM might be heated to $T \sim 100
\unit{K}$ at $z < 10$ and ionized to nearly 0.5\%. The interaction
between higher-energy X-rays and the IGM is too weak to have a non-negligible
effect on the thermal and ionization state of the IGM.  
When taking the spectrum energy distribution of the Pop III binary 
X-rays into account, the heating and ionization effects are weaker than the 1 keV case, 
but they are still substantial. 
The IGM heating might be detectable through 21 cm
observations by SKA at $z \la 15$.  Since the temperature profiles are so
different for different X-ray photon energies, 21 cm observations
might possibly constrain the X-ray SED from Pop III
binaries. The details on the possible 21-cm observations of these
heated IGM will be reported in a forthcoming paper \citep{Ahn13}. On
the other hand, the increased ionization is so weak that it does not
significantly contribute to the optical depth of the CMB to Thomson
scattering. The ionization at these redshifts might then be dominated by
the UV radiation from Pop II stars in low-mass metal cooling halos \citep{Wise14}, 
which just form following the SN explosion of Pop III stars considered here.

We find that the sub-keV photons are most effective in locally heating
and ionizing the IGM, and their effects on the IGM are much stronger than
those by the X-ray background of higher-energy photons. Our simulated
Rarepeak is a 3.5$\sigma$ density peak and the average distance between two
similar peaks is only $\sim$ 100 comoving Mpc, suggesting that the IGM
is more sensitive to lower-energy X-rays from nearby sources than to
X-ray backgrounds of higher-energy photons.

The sources and amount of X-ray radiation in the early universe are
under debate. Here, we only consider the possibility of Pop III
binaries, but ignore other major sources, such as quasars,
mini-quasars, and supernova remnants.  However, our work suggests that the
X-rays from Pop III binaries might dominate all other sources, at
least at $z > 10$.  It is generally thought that the two most important
sources of X-rays in the universe are AGNs and XRBs (non-Pop III).  The X-ray
luminosity density inside Rarepeak\footnote{Even in an unlikely
  scenario that Rarepeak contains the only X-ray sources inside the
  simulated volume, the X-ray luminosity density over the entire
  simulated box is still $\sim$ 10$^{38}$ erg s$^{-1}$ per comoving
  Mpc$^{3}$.} is $\sim$ 5 $\times$ 10$^{40}$ erg s$^{-1}$ per comoving
Mpc$^{3}$, which is $\sim$ 10 times higher than that from normal XRBs at
their peak at $z \sim 3$, and 10$^3$ times higher than that at z
$\sim$ 15 \citep{Fragos13}.  X-rays from AGNs, estimated from both
observations \citep{Hasinger05,Hopkins07,Silverman08,Aird10} and
semi-analytic models with N-body simulations \citep{Croton06}, are
only substantial at $z < 6$, and even at their peak at z $\sim$ 3,
their X-ray luminosity density is $<$ 10$^{40}$ erg s$^{-1}$
comoving Mpc$^{-3}$.  

However, since the Pop III initial mass function, binary fraction 
and evolution are not yet well constrained, our estimate 
of the X-ray luminosities and Pop III binary SED are alone very uncertain, 
depending on the choice of model parameters, for instance, Pop III binary 
occurrence and BH mass.   We might 
overestimate the X-ray output by assuming high Pop III
binary occurrence and/or long binary lifetime by a factor of a
few. However, even in such a case, Pop III binaries should still be the major
sources of X-rays at high redshifts. On the other hand, we did not
consider any X-rays from the later accretion to the BH after
the companion of the binary dies, which could lead to an
underestimation of the early X-ray background. However, we confirmed
in the upper limit of the binaries accreting indefinitely that the Pop
III contribution to the X-ray background is only a factor of a few higher.
Our calculations show that even if the X-ray flux is 1--2 orders of
magnitude stronger, it is still within the linear regime, and our
results can be easily adjusted to another X-ray background
that includes a more accurate model of Pop III binaries and other
X-ray sources.
  
We have not considered the effects of relative streaming velocities
(v$_{vel}$ $\sim$ 30 km s$^{-1}$ at z $\sim$ 1100) between baryons and dark matter
that arise during recombination \citep{Tseliakhovich10}.
This phenomenon only suppresses Pop III star formation in the
smallest mini-halos with M $\le$ 10$^6$ M$_\odot$ \citep{Tseliakhovich11, Greif11, 
 Naoz12,OLeary12}, which are not well resolved in our simulation, and should not 
 significantly change our results. Additionally, after both Pop III and Pop II star formation 
 gets going, the minimum halo mass in which Pop III star formation takes place rises to be substantially 
 over 10$^6$ M$_\odot$ due to the LW background \citep{Crosby13a}, thus making 
 this effect quite unimportant. \citet{Xu13} actually showed that the Pop III star
 formation in a few 10$^6$ M$_\odot$ halos in the high-density region 
 of Rarepeak is already suppressed by the 
 LW radiation by redshift z $\sim$ 18. For studying an earlier phase of Pop III 
 X-ray evolution, however, one may still need to increase mass resolution 
 and include the relative streaming velocities simultaneously.

Currently, it is computationally unfeasible to execute a full
radiation hydrodynamic simulation to $z=6$ in such a large high-density volume
that resolves all Pop III-forming halos and calculates their formation rate.  
Without the exact evolution and distribution of Pop III formation at lower redshifts, we can only use a
time-independent X-ray distribution to continue our calculations from
$z=15$.  This assumption might not be much
different than a self-consistent calculation of the X-ray background
from Pop III binaries, though.  As discussed in \citet{Xu13}, we
expect that the Pop III formation will continue in the Rarepeak region
but will be gradually suppressed.  It is also likely that the Pop III
formation in the other lower-density regions will become comparable to
overdense regions like the Rarepeak.  Recall that the halo mass
function of this region at $z=15$ is similar to the $z=10$ halo mass
function of an average patch of the universe.  Thus, it is reasonable
to expect that the X-rays from Pop III binaries will continue to much
lower redshifts  z $\sim$ 10 \citep[also see][]{Crosby13a}.  We are currently 
running a similar simulation of an average region to lower redshifts. This, complementing 
the Rarepeak simulation, will provide a more complete understanding of the 
Pop III formation and X-ray background histories before the end of the reionization.


\acknowledgements 

We thank Jordan Mirocha for his comments on the paper. This research 
was supported by National Science Foundation (NSF) grant
AST-1109243 to MLN. KA was supported by the NRF grant funded by the Korean 
government MEST (No. 2012R1A1A1014646). KA also acknowledges a very generous 
support from M. Norman for KA's sabbatical visit to UCSD, where part of this work was 
performed as a collaboration between KA and Norman's group members. JHW acknowledges support from NSF grants
AST-1211626 and AST-1333360.  BWO was supported in part by the MSU 
Institute for Cyber-Enabled Research. The simulation was performed using \texttt{Enzo} on the
Kraken supercomputer operated for the Extreme Science and Engineering
Discovery Environment (XSEDE) by the National Institute for
Computational Science, ORNL with XRAC allocation MCA-TG98020N, and on the
Blue Waters operated by the National Center for Supercomputing Applications (NCSA) with 
PRAC allocation support by the NSF (award number OCI-0832662). Data
analysis was performed on the Gordon supercomputer operated for XSEDE by
the San Diego Supercomputer Center and on the Blue Waters supercomputer. This research is part of the Blue Waters 
sustained-petascale computing project, which is supported by the NSF
(award number ACI 1238993) and the state of Illinois. Blue Waters is a joint effort of the University 
of Illinois at Urbana-Champaign and its NCSA.  
This research has made use of NASA’s Astrophysics Data System Bibliographic
Services.  The majority of the analysis and plots were done with
\texttt{yt} \citep{yt_full_paper}.
\texttt{Enzo} and \texttt{yt} are developed by a large number of independent researchers 
from numerous institutions around the world. Their commitment to open science has helped make this work possible.


\end{document}